\documentclass[preprint,5p,times,twocolumn]{elsarticle}
\usepackage{graphics}
\usepackage{subfigure}
\usepackage{amssymb}
\usepackage{amsmath}
\usepackage{mathptmx}
\usepackage{etoolbox}
\usepackage{mathtools}
\usepackage[dvipsnames]{xcolor}
\journal{Submitted to Physica D: Nonlinear Phenomena}
\begin{document}
%%%%%%%%%%%%%%%%%%%%%%%%%%%%%%%%%%%%%%%%%%%%%%%%%%%%%%%%%%%%%%%%%%%%%%%%%%%%%%%%
\begin{frontmatter}
%-------------------------------------------------------------------------------
\title{Nonlinear $q$-voter model involving nonconformity on networks}
%-------------------------------------------------------------------------------
\author[ugm]{Rinto Anugraha NQZ \corref{cp}}
\ead{rinto@ugm.ac.id}
\cortext[cp]{Corresponding author}

%-------------------------------------------------------------------------------

\author[apctp,brin]{Roni Muslim}
\ead{roni.muslim@apctp.org, roni.muslim@brin.go.id}
% \ead{muslim.roni@gmail.com}

%-------------------------------------------------------------------------------
\author[itk]{Henokh Lugo H.}%
\ead{henokh.lugo@lecturer.itk.ac.id}

%-------------------------------------------------------------------------------

\author[ugm]{Fahrudin Nugroho}
\ead{fakhrud@ugm.ac.id}

\author[ugm]{Idham Syah Alam}
\ead{idham@ugm.ac.id}

\author[uin]{Muhammad Ardhi Khalif}
\ead{muhammad\_ardhi@walisongo.ac.id}

\affiliation[ugm]{
    organization={Department of Physics,
    Universitas Gadjah Mada},
    city={Yogyakarta},
    postcode={55281},
    country={Indonesia}
}

\affiliation[apctp]{
    organization={Asia Pacific Center for Theoretical Physics (APCTP)},
    city={Pohang},
    postcode={37673},
    country={South Korea}
}

\affiliation[brin]{
    organization={Research Center for Quantum Physics,
    National Research and Innovation Agency (BRIN)},
    city={South Tangerang},
    postcode={15314},
    country={Indonesia}
}

\affiliation[itk]{
    organization={Information System,
     Institut Teknologi Kalimantan},
    city={Balikpapan},
    postcode={76127},
    country={Indonesia}
}
\affiliation[uin]{
    organization={Department of Physics, Walisongo State Islamic University},
    city={Semarang},
    postcode={50181},
    country={Indonesia}
}

%-------------------------------------------------------------------------------
\begin{abstract}
The order-disorder phase transition is a fascinating phenomenon in opinion dynamics models within sociophysics. This transition emerges due to noise parameters, interpreted as social behaviors such as anticonformity and independence (nonconformity) in a social context. In this study, we examine the impact of nonconformist behaviors on the macroscopic states of the system. Both anticonformity and independence are parameterized by a probability \( p \), with the model implemented on a complete graph and a scale-free network. Furthermore, we introduce a skepticism parameter \( s \), which quantifies a voter's propensity for nonconformity. Our analytical and simulation results reveal that the model exhibits continuous and discontinuous phase transitions for nonzero values of \( s \) at specific values of \( q \). We estimate the critical exponents using finite-size scaling analysis to classify the model's universality. The findings suggest that the model on the complete graph and the scale-free network share the same universality class as the mean-field Ising model. Additionally, we explore the scaling behavior associated with variations in \( s \) and assess the influence of \( p \) and \( s \) on the system's opinion dynamics.
\end{abstract}
%-------------------------------------------------------------------------------
%%Graphical abstract
%\begin{graphicalabstract}
%\centering
%\includegraphics[width=16cm]{grabs}
%\end{graphicalabstract}

%Research highlights
% \begin{highlights}
% \item Analysis of the dynamics of the $q$-voter model, involving two distinct social behaviors and skepticism on networks.
% \item Investigation of phase transitions and the emergence of scaling phenomena.
% \item Examination of the model's universality class through finite-size scaling analysis.
% \end{highlights}
%-------------------------------------------------------------------------------
%-------------------------------------------------------------------------------
\end{frontmatter}
%%%%%%%%%%%%%%%%%%%%%%%%%%%%%%%%%%%%%%%%%%%%%%%%%%%%%%%%%%%%%%%%%%%%%%%%%%%%%%%%
\section{\label{sec:Sec1}Introduction}
%-------------------------------------------------------------------------------
Over the past few decades, scholars have increasingly applied statistical physics's general concepts and principles to various interdisciplinary fields, including socio-political studies, economics, biology, medicine, computer science, and more. In particular, applying statistical physics to understanding socio-political phenomena has led to the development of several models of opinion dynamics. These models aim to describe the interaction rules among individuals that influence collective agreement or dissent. The development of opinion dynamics models does not follow a universal standard; they can arise from direct observation of real-world social phenomena or reasonable theoretical assumptions. For further insights into the perspectives of scientists regarding the application of statistical physics to social phenomena, readers are referred to Refs.~\cite{galam2012socio, castellano2009statistical, sen2014sociophysics, schweitzer2018sociophysics}.

In addition to analyzing real socio-political phenomena, researchers have sought to deepen their understanding of physical phenomena arising in opinion dynamics, such as the order-disorder phase transition in the Ising model. In the context of opinion dynamics, this order-disorder phase transition is often referred to as the transition between consensus and status quo, occurring at a critical value of a social parameter. This phase transition is one of the most intriguing physical phenomena to explore within opinion dynamics models, as it provides insight into macroscopic changes in a system driven by complex microscopic interactions. Such transitions are typically triggered by small fluctuations in system parameters, analogous to the ferromagnetic-antiferromagnetic transition observed in magnetic spin systems due to temperature changes. This phenomenon has also been extensively studied in sociophysics models of opinion dynamics \cite{holyst2000phase,li2012phase,mukherjee2016disorder,velasquez2017interacting,muslim2020phase,muslim2021phase,schawe2022higher,MUSLIM2022133379,MUSLIM2022128307}.  

Beyond phase transitions induced by small changes in noise parameters, researchers have also investigated scaling behavior and universality in opinion dynamics. Scaling refers to the data collapse on a particular parameter, often termed parameter scaling. Similarly, the universality of a model is defined by specific critical parameters that cause the macroscopic behavior of the system to align. For instance, the universality class of mean-field models corresponds to those defined on complete graphs (fully connected networks) \cite{muslim2020phase, muslim2021phase, MUSLIM2022133379, MUSLIM2022128307}. Meanwhile, opinion dynamics models defined on two- and three-dimensional networks exhibit universality classes that align with the Ising models in corresponding dimensions \cite{calvelli2019phase, muslim2024impact}, as described in classical studies of phase transitions and critical phenomena \cite{stanley1971phase, campostrini2002critical}. 

In the context of social systems, various social parameters—such as conformity, anticonformity, and independence—are employed to observe and interpret social phenomena through principles derived from physics. In the social science literature, these parameters are typically classified into two categories: conformity and nonconformity \cite{willis1963two, willis1965conformity, hofstede2010}. These concepts have been widely studied within the framework of opinion dynamics models in recent years \cite{muslim2020phase, muslim2021phase, MUSLIM2022133379, MUSLIM2022128307, grabisch2020survey, baron2021consensus, lipiecki2022polarization, azhari2023independence, mulya2024phase, fardela2024opinion, muslim2024impact}. As highlighted earlier, these social parameters not only enrich the model with features characteristic of statistical physics but also make it more applicable to real-world social phenomena \cite{oestereich2020hysteresis, civitarese2021external, doniec2022consensus, sznajd2024toward}.  

This paper extends the opinion dynamics model previously discussed in Ref.~\cite{civitarese2021external} by incorporating social anticonformist behavior represented through a probability parameter \( p \). In Ref.~\cite{civitarese2021external}, the authors introduced an external field, denoted by \( m \), to model the influence of mass media in a social system, along with a skepticism parameter \( s \), which represents an agent's reluctance to conform. Their findings demonstrated that the interplay between these parameters can induce phase transitions in the system.  

Several studies have explored analogous scenarios, particularly the effects of an external field—representing mass media influence—on the macroscopic state of the system, including shifts in consensus behavior \cite{candia2008mass, martins2010mass, crokidakis2012effects, azhari2023external, MUSLIM2024129358, fardela2024opinion}. This concept of an external field closely resembles its role in the Ising model, where it alters the critical point and drives the system through an order-disorder phase transition \cite{muslim2020phase, muslim2021phase, MUSLIM2022133379, MUSLIM2022128307, nyczka2012phase, vieira2016phase, calvelli2019phase}. As noted in Ref.~\cite{civitarese2021external}, the external field exerts a uniform influence across all agents in the system, enabling agents to collectively shift their opinions, such as transitioning from an ``up" state to a ``down" state (or vice versa), under the field's effect.

We build upon the scenario proposed in Ref.~\cite{civitarese2021external}, introducing an additional social response parameter representing anticonformity. The model is implemented on two types of networks: a complete graph and a scale-free network, allowing for a detailed analysis of how the parameters \( p \) (the probability of anticonformity) and \( s \) (the skepticism level) influence the macroscopic state of the system. We also examine the effects of these parameters on the system's opinion dynamics.  

Our findings reveal distinct patterns that diverge from those reported in the independence-based model studied in Ref.~\cite{civitarese2021external}. Furthermore, we extend the analysis to include additional statistical properties, such as scaling behavior and the universality class of the model, which were not explored in the previous study. This comprehensive investigation enhances our understanding of the model's dynamic behavior and critical properties, particularly under varying network structures and social response parameters.

%%%%%%%%%%%%%%%%%%%%%%%%%%%%%%%%%%%%%%%%%%%%%%%%%%%%%%%%%%%%%%%%%%%%%%%%%%%%%%%%
\section{\label{sec:Sec2} Model Description}
%-------------------------------------------------------------------------------
The nonlinear \( q \)-voter model,- an extension of the original voter model, was introduced by Castellano et al.~\cite{castellano2009nonlinear}. This model incorporates a nonlinear flip probability function, \( f(r,q) = r^q + \varepsilon \left[ 1 - r^q - \left( 1 - r \right)^q \right] \), where \( r \) represents the fraction of disagreeing neighbors, and \( \varepsilon \) denotes the probability of an agent changing its opinion when its \( q \)-neighbors are not in a homogeneous state. For \( q = 1 \) or \( q = 2 \) with \( \varepsilon = \frac{1}{2} \), the function simplifies to a linear form, \( f(r) = r \), which corresponds to the original voter model. The nonlinear behavior emerges for \( q \geq 2 \) and \( \varepsilon \neq \frac{1}{2} \).

The parameter \( \varepsilon \) in the flip function \( f(r, q) \) acts as noise, disrupting the ordered state of the model. While its interpretation can vary, in a social context, it often corresponds to psychological behaviors such as independence or anticonformity (also referred to as contrarianism), represented by a probability parameter \( p \). Anticonformity opposes the majority opinion, whereas independence reflects resistance to external influences, characterized by autonomous decision-making \cite{willis1963two, willis1965conformity}. Previous studies have identified a relationship between nonconformity and Hofstede's individualism index, although the two concepts are not directly equivalent. Societies with high individualism index scores tend to exhibit these behaviors more prominently \cite{hofstede2001, hofstede2010, solomon2010consumer}. For more details on the individualism index, see Ref.~\cite{hofstedeIDV}.

As noted earlier, our model incorporates independence and anticonformity behaviors through two parameters: \( p \in [0,1] \), representing the probability of nonconformity, and \( s \in [0,1] \), denoting the agents' susceptibility to such behaviors. A higher value of \( s \) indicates a stronger inclination toward nonconformity, reflecting greater skepticism and a reduced likelihood of being influenced by peers or external forces. The skepticism parameter \( s \) in our model functions analogously to the skepticism parameter introduced in Ref.~\cite{civitarese2021external}, as both measure agents' resistance to change their opinions.  

The key distinction lies in the interpretation of \( s \). In our model, \( s \) encapsulates a generalized tendency toward nonconformity, signifying a broad reluctance to conform to social norms or majority opinions. By contrast, in Ref.~\cite{civitarese2021external}, \( s \) specifically represents skepticism directed toward an external field, such as media influence. Higher \( s \) values in their model indicate a reduced likelihood of agents being swayed by media-driven majority opinions, making them more inclined to consider alternative viewpoints. Thus, while both models utilize \( s \) to capture nonconformity, our approach generalizes \( s \) as a universal resistance to social conformity, whereas in Ref.~\cite{civitarese2021external}, \( s \) highlights agents' distrust of centralized information sources and its implications for consensus dynamics.

In the context of these behaviors, the model operates as follows: Starting from an initial random state where the opinion density is \( 1/2 \), a group of agents is randomly selected to form a committee of size \( q \) (referred to as the \( q \)-sized agent), and an additional voter aligns their opinion with the \( q \)-sized agent. A random number \( r \) is generated to determine the voter's behavior.  If \( r \leq p \), the voter acts independently with a probability \( p \). Additionally, with a probability \( s \), the voter behaves skeptically, leading to a \( 1/2 \) probability of changing their opinion. Consequently, the probability of an opinion change—from opinion-up to opinion-down or vice versa—is \( ps/2 \). If \( r > p \), the voter follows the standard \( q \)-voter model rule. In this case, with a probability of \( (1 - p) \), the voter adopts the opinion of the $q$-sized agent when the $q$-sized agent shares the same opinion, reflecting conformity behavior. The model is illustrated in~\eqref{eq:illust_indep}:
\begin{equation}\label{eq:illust_indep}
    \begin{aligned}
        & \text{independence} & \cdots \quad  &\Uparrow \quad \stackrel{}{\rightarrow}\quad \cdots \quad & \Downarrow, \\
        & \text{conformity} &\, \downarrow \downarrow  \downarrow  \quad &\Uparrow \quad \stackrel{}{\rightarrow} \quad \downarrow \downarrow \downarrow \quad & \Downarrow, \\
    \end{aligned}
\end{equation}
where \( \uparrow \) (\( \downarrow \)) denotes an opinion-up (opinion-down) among the members of the \( q \)-sized agent, while \( \Uparrow \) (\( \Downarrow \)) represents the opinion-up (opinion-down) of the voter. The conformity scenario illustrated in~\eqref{eq:illust_indep} corresponds to the case of \( q = 3 \).

In scenarios involving anticonformity behavior, if the voter and the \( q \)-sized agent initially share the same opinion, the voter changes their opinion to the opposite of the \( q \)-sized agent. Conversely, if the voter and the \( q \)-sized agent have differing opinions, the voter follows the standard \( q \)-voter model rule. This behavior is illustrated in \eqref{eq:illust_anti}:
\begin{equation}\label{eq:illust_anti}
    \begin{aligned}
        & \text{anticonformity} &\, \uparrow \uparrow \uparrow  \quad  &\Uparrow \quad \stackrel{}{\rightarrow} \quad &\uparrow \uparrow \uparrow \quad &\Downarrow, \\
        & \text{conformity} &\, \uparrow \uparrow \uparrow \quad &\Downarrow \quad \stackrel{}{\rightarrow}\quad &\uparrow \uparrow \uparrow \quad &\Uparrow. \\
    \end{aligned}
\end{equation}

\begin{figure}[t]
    \centering
    \includegraphics[width=0.5\linewidth]{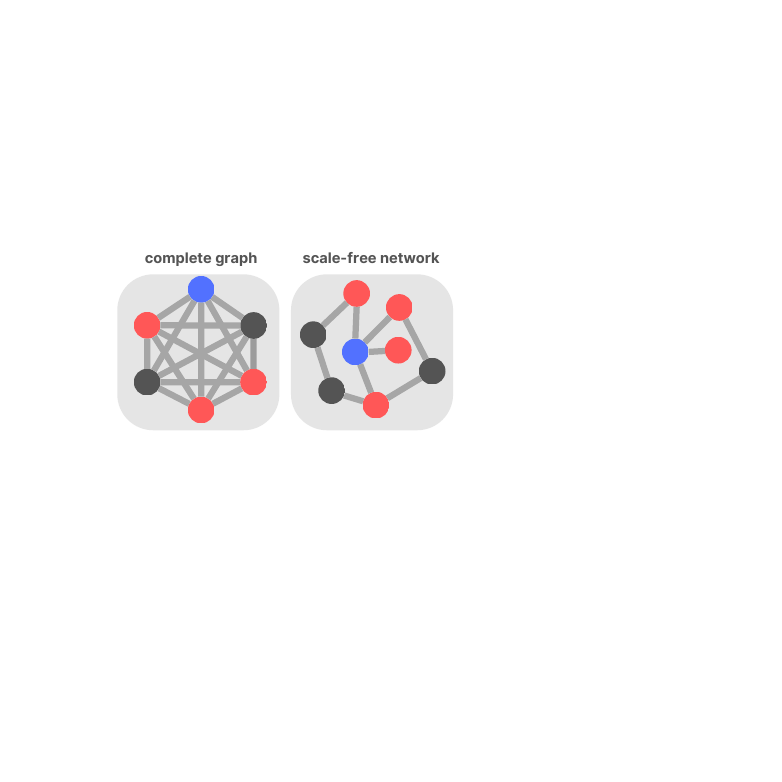}
    \caption{The two network topologies considered in the model: the complete graph and the scale-free network. Blue and red nodes represent voters and randomly selected \( q \)-sized agents, respectively. Black nodes represent agents not selected within the population.}
    \label{fig:network}
\end{figure}

We examined the model on two types of networks: the complete graph and the scale-free network (Barabási-Albert network), as illustrated in Fig.~\ref{fig:network}. In the scale-free network, the \( q \)-sized group of nearest neighbors is randomly selected from the available nearest neighbors. In this setup, each node in the network has at least \( q \) nearest neighbors. From a social perspective, the scale-free network provides a more realistic representation of social connections than the complete graph, which assumes homogeneous interactions across all agents \cite{newman2018networks}.  

The nonlinear \( q \)-voter model with external field effects, as explored in previous work \cite{civitarese2021external}, shares similarities with the independence type examined in this study. Here, we analyze the model analytically and numerically for two types of social interactions: independence and anticonformity, represented by the probability \( p \). The analysis is conducted on the complete graph (mean-field approximation) and the scale-free network, focusing on identifying the critical point and exponents using finite-size scaling analysis.  

The finite-size scaling relations are defined as follows:
$m = \phi_{m}(x) N^{-\beta/\nu}, \quad \chi = \phi_{\chi}(x) N^{\gamma/\nu}, \quad p - p_{c} \sim  N^{-1/\nu}, \quad \text{and} \quad U = \phi_{U}(x)$,
where \( U = 1 - \langle m^4 \rangle/(3 \langle m^2 \rangle^2) \) is the fourth-order Binder cumulant, and \( \chi = N ( \langle m^2 \rangle - \langle m \rangle^2 ) \) represents the susceptibility \cite{binder2012monte}. The parameter \( \phi_i(x) \) denotes the dimensionless scaling function that characterizes finite-size scaling effects, with \( x = (p - p_c) N^{1/\nu} \) and \( i = \{m, \chi, U\} \).

The order parameter \( m \) (magnetization) can also be computed using
\begin{equation}\label{eq:Eq3}
    m =  \left|\dfrac{1}{N} \sum \sigma_i \right| = \left|\dfrac{N_{\uparrow}-N_{\downarrow}}{N_{\uparrow}+N_{\downarrow}} \right|,
\end{equation}
where each agent holds one of two possible opinions, denoted by \( \sigma_i = \pm 1 \), corresponding to Ising spins with \(\uparrow = +1\) and \(\downarrow = -1\). This analysis includes an examination of the impact of skepticism (\( s \)) on the outcome related to the anticonformity probability (\( p \)), which is a novel contribution of this study and is not covered in the previous work~\cite{civitarese2021external}.

%%%%%%%%%%%%%%%%%%%%%%%%%%%%%%%%%%%%%%%%%%%%%%%%%%%%%%%%%%%%%%%%%%%%%%%%%%%%%%%%
\section{\label{sec:Sec3} Result and Discussion}
%-------------------------------------------------------------------------------
\subsection{Mean-field approximation}
The mean-field approximation assumes a homogeneous and isotropic system, neglecting all fluctuations within the system. This simplification equates local and global concentrations, describing the system using a single parameter, such as the fraction of opinions \( k \) in this model \cite{binney1992theory, amit2005field}. The mean-field approach is independent of spatial dimensions, further simplifying the system's analysis. In terms of graph structure, this approximation reduces the complexity of the graph, enabling more tractable analytical treatments.  Despite these simplifications, the mean-field approximation retains the capacity to capture key features of statistical physics, such as phase transitions (both continuous and discontinuous), scaling behavior, and universality. These phenomena have been the focus of numerous recent studies \cite{MUSLIM2022133379, MUSLIM2022128307,   muslim2024impact, mulya2024phase, MUSLIM2024129358, mili2022simple, rocha2021novel, li2020some}. In this model, we observe that the system exhibits phase transitions and scaling behavior driven by two noise-like parameters: independence and anticonformity.

\subsubsection{Model with independence}

In this model, an independent voter acts autonomously within the population, changing its opinion without being influenced by neighboring agents. The dynamics of this autonomous behavior are determined by the probability \( p \), which represents the likelihood of an agent acting independently, and \( s \), which quantifies the agent's skepticism toward others' opinions. Specifically, an agent behaves independently with probability \( p \) and exhibits skepticism with probability \( s \). When a voter considers changing their opinion, there is a \( 1/2 \) probability that the voter will reverse their current opinion, expressed as \( \pm S_i(t) = \mp S_i(t+1) \). Consequently, the overall probability of a voter changing his/her opinion is given by \( ps/2 \). 

We follow the methodology outlined in Ref.~\cite{civitarese2021external} to describe the model's dynamics. Specifically, we define the probabilities for an increase (\(\varphi_{+}\)) and a decrease (\(\varphi_{-}\)) in the fraction of opinion \( \uparrow \) when \( N_{\uparrow}/N \leq 1/2 \) in a finite system as follows:
\begin{align}
\varphi_{+} & = N_{\downarrow}\left[ \dfrac{\left(1-p\right) \prod_{j =1}^{q} \left(N_{\uparrow}-j+1\right)}{\prod_{j=1}^{q+1}\left(N-j+1\right)} +\dfrac{p\,s}{2\,N} \right], \label{eq:Eq4}\\
\varphi_{-} & = N_{\uparrow}\left[ \dfrac{\left(1-p\right) \prod_{j =1}^{q} \left(N_{\downarrow}-j+1\right)}{\prod_{j=1}^{q+1}\left(N-j+1\right)} +\dfrac{p}{N} \left(1-\dfrac{s}{2}\right) \right], \label{eq:Eq5}
\end{align}
and for $N_{\uparrow}/N > 1/2$:
\begin{align}
\varphi_{+} & = N_{\downarrow}\left[ \dfrac{\left(1-p\right) \prod_{j =1}^{q} \left(N_{\uparrow}-j+1\right)}{\prod_{j=1}^{q+1}\left(N-j+1\right)}+ \dfrac{p}{N} \left(1-\dfrac{s}{2}\right)\right], \label{eq:Eq6}\\
\varphi_{-} & = N_{\uparrow}\left[ \dfrac{\left(1-p\right) \prod_{j =1}^{q} \left(N_{\downarrow}-j+1\right)}{\prod_{j=1}^{q+1}\left(N-j+1\right)} +\dfrac{p\,s}{2\,N} \right] \label{eq:Eq7}.
\end{align}

Equations \eqref{eq:Eq4}–\eqref{eq:Eq7} describe the dynamics of the system, specifying the probabilities of increasing or decreasing the fraction of opinion density by \( +1/N \) or \( -1/N \), respectively, or remaining unchanged with a probability \( (1 - \varphi_{+} - \varphi_{-}) \), for a specific range of the fraction of opinion density \( N_{\uparrow}/N \).  To evaluate the accuracy of the analytical predictions compared to Monte Carlo (MC) simulations, we can expand Eqs.~\eqref{eq:Eq4}–\eqref{eq:Eq7} for arbitrary \( N \). However, the mean-field approximation becomes increasingly valid for large \( N \gg 1 \). Therefore, our primary interest lies in examining Eqs.~\eqref{eq:Eq4}-\eqref{eq:Eq7} as:
\begin{align}
   &\varphi_+  \approx  
  \begin{dcases} \label{eq:Eq8}
    \left(1-k\right)\left[\left(1-p\right)k^q + \frac{ps}{2} \right] & \text{for} \,k \leq \dfrac{1}{2}, \\
  \left(1-k\right)\left[\left(1-p\right)k^q + p \left(1-\frac{s}{2}\right)\right] & \text{for} \,k > \dfrac{1}{2},
  \end{dcases} \\
   &\varphi_{-} \approx  
  \begin{dcases} \label{eq:Eq9}
    k\left[\left(1-p\right)\left(1-k\right)^q+ p \left(1-\frac{s}{2}\right)\right] & \text{for} \,k \leq \dfrac{1}{2}, \\
   k\left[\left(1-p\right)\left(1-k\right)^q+ \frac{ps}{2}\right] & \text{for} \,k > \dfrac{1}{2},
  \end{dcases} 
\end{align}
where $k = N_{\uparrow}/N$. 

\subsubsection{Model with anticonformity}
In this model, the voter exhibits anticonformist behavior with probability \( p \), and with probability \( s \), the voter expresses skepticism toward the opinion of the \( q \)-sized agent. When the \( q \)-sized agent holds a unanimous opinion, the voter adopts the opposite opinion, expressed as \(\pm S_i(t) = \mp S_i(t+1)\). As in the model with independence, the probabilities for a fraction of opinions to increase (\(\varphi_{+}\)) or decrease (\(\varphi_{-}\)) in the finite system, for \( N_{\uparrow}/N \leq 1/2 \), can be expressed as follows: 
\begin{align}
 \varphi_{+} & = \dfrac{\left(1-p \right) N_{\downarrow} \prod_{j = 1}^{q} \left(N_{\uparrow}-j+1 \right)}{\prod_{j = 1}^{q+1} \left(N-j+1 \right)} + \dfrac{p\,s \prod_{j = 1}^{q+1} \left(N_{\downarrow}-j+1 \right)}{\prod_{j = 1}^{q+1} \left(N-j+1 \right)}, \label{eq:Eq10}\\
 \varphi_{-} & = \dfrac{\left(1-p \right) N_{\uparrow} \prod_{j = 1}^{q} \left(N_{\downarrow}-j+1 \right)}{\prod_{j = 1}^{q+1} \left(N-j+1 \right)} + \dfrac{p\prod_{j = 1}^{q+1} \left(N_{\uparrow}-j+1 \right)}{\prod_{j = 1}^{q+1} \left(N-j+1 \right)},\label{eq:Eq11}
\end{align}
and for $N_{\uparrow}/N > 1/2$:
\begin{align}
 \varphi_{+} & = \dfrac{\left(1-p \right) N_{\downarrow} \prod_{j = 1}^{q} \left(N_{\uparrow}-j+1 \right)}{\prod_{j = 1}^{q+1} \left(N-j+1 \right)} + \dfrac{p\prod_{j = 1}^{q+1} \left(N_{\downarrow}-j+1 \right)}{\prod_{j = 1}^{q+1} \left(N-j+1 \right)}, \label{eq:Eq12} \\
 \varphi_{-} & = \dfrac{\left(1-p \right) N_{\uparrow} \prod_{j = 1}^{q} \left(N_{\downarrow}-j+1 \right)}{\prod_{j = 1}^{q+1} \left(N-j+1 \right)} + \dfrac{p\,s \prod_{j = 1}^{q+1}, \left(N_{\uparrow}-j+1 \right)}{\prod_{j = 1}^{q+1} \left(N-j+1 \right)}\label{eq:Eq13}.
\end{align}
Again, for comparison with numerical simulations on the complete graph, we analyze Eqs.~\eqref{eq:Eq10}–\eqref{eq:Eq13} in the limit of a large population size \( N \). In this regime, the equations can be simplified to: 
\begin{align}
   &\varphi_+  \approx  
  \begin{dcases} \label{eq:Eq14}
    \left(1-k\right)\left[\left(1-p\right)k^q + ps(1-k)^{q}\right] & \text{for} \, k \leq \dfrac{1}{2}, \\
      \left(1-k\right)\left[\left(1-p\right)k^q + p\left(1-k\right)^{q} \right] & \text{for} \, k > \dfrac{1}{2},
  \end{dcases} \\
   &\varphi_{-} \approx  
  \begin{dcases} \label{eq:Eq15}
    k\left[\left(1-p\right)\left(1-k\right)^q+ p\,k^{q}\right] & \text{for} \, k \leq \frac{1}{2}, \\
   k\left[\left(1-p\right)\left(1-k\right)^q+ p\,s\,k^{q}\right] & \text{for} \, k > \frac{1}{2}.
  \end{dcases} 
\end{align}

\subsection{\label{subsec:Subsec.3.2} Time evolution and steady state}
The recursive formula for the fraction of opinions \( k \) at time \( t \) can be derived from the discrete-time master equation. For a finite population size \( N \), the evolution of \( k \) is described by the following equation \cite{krapivsky2010kinetic}:  
\begin{align} \label{eq:Eq17}
    k(t+1) = k(t) + \dfrac{1}{N} \left[\varphi_{+}(k) - \varphi_{-}(k) \right],
\end{align}  
where \(\varphi_{+}(k)\) and \(\varphi_{-}(k)\) represent the probabilities of \( k \) increasing and decreasing during the dynamical process, as defined in Eqs.~\eqref{eq:Eq8}–\eqref{eq:Eq9} and Eqs.~\eqref{eq:Eq14}–\eqref{eq:Eq15}.  The time step \(\Delta t = 1/N\) corresponds to one MC step. For a large population size (\( N \to \infty \)) or in the limit where \(\Delta t \to 0\), Eq.~\eqref{eq:Eq17} can be expressed in its continuous form as:  
\begin{equation}\label{eq:Eq18}
    \dfrac{\mathrm{d}k(t)}{\mathrm{d}t} = \varphi_{+}(k)-\varphi_{-}(k).
\end{equation}
Equation~\eqref{eq:Eq18} is commonly referred to as the rate equation for the fraction of opinions \( k \), which describes the time evolution of opinions within the system.  

The macroscopic behavior of the system can be analyzed by examining the equilibrium condition of Eq.~\eqref{eq:Eq18}, where \(\mathrm{d}k(t)/\mathrm{d}t = 0\). This condition implies that \(\varphi_+(k) = \varphi_-(k)\). By substituting this equilibrium condition into Eqs.~\eqref{eq:Eq8}–\eqref{eq:Eq9} and Eq.~\eqref{eq:Eq18}, the probability of independence \( p \) in the stationary state of the model can be determined as: 
\begin{align}\label{eq:Eq19}
   p  =  
  \begin{dcases} 
    \dfrac{k_{st}\left(1-k_{st}\right)^q-k_{st}^q\left(1-k_{st}\right)}{k_{st}\left(1-k_{st}\right)^q-k_{st}^q\left(1-k_{st}\right)-k_{st}+s/2} & \text{for} \, k \leq \dfrac{1}{2}, \\
  \dfrac{k_{st}\left(1-k_{st}\right)^q-k_{st}^q\left(1-k_{st}\right)}{k_{st}\left(1-k_{st}\right)^q-k_{st}^q\left(1-k_{st}\right)-k_{st}+1-s/2} & \text{for} \, k > \dfrac{1}{2}.
  \end{dcases}
\end{align}
Equation \eqref{eq:Eq19} describes the relationship between the probability \( p \) and skepticism \( s \) at equilibrium for specific ranges of \( k \leq 1/2 \) and \( k > 1/2 \). The relationship becomes equivalent when \( s = 1 \). If we plot \( s \) as a function of \( p \) for a specific \( k \), we observe distinct behaviors: skepticism \( s \) exhibits a non-smooth decrease at low \( p \) values and for large \( q \)-sized agents, while it shows a smooth decrease for small \( q \)-sized agents. For all values of \( q \), \( k \) increases as \( s \) and \( p \) increase as discussed in Ref.~\cite{civitarese2021external}.  

An order-disorder phase transition can also be identified within this model by analyzing the equilibrium condition of Eqs.~\eqref{eq:Eq8}–\eqref{eq:Eq9}. Consequently, the probability of independence \( p \) at equilibrium is expressed as:
\begin{equation} \label{eq:Eq20}
    p = \left[\dfrac{1}{2}\left(s\left(1-2\,k_{st}\right)\left[k_{st}\left(1-k_{st}\right)^q-k_{st}^q\left(1-k_{st}\right)\right]^{-1}\right)+1\right]^{-1}.
\end{equation}
It is important to note that Eqs.~\eqref{eq:Eq19} and \eqref{eq:Eq20} differ slightly in their scope. Equation~\eqref{eq:Eq19} describes the phase diagram for a specific range of \( k \), whereas Eq.~\eqref{eq:Eq20} extends to cover the phase diagram for all values of \( k \). The critical independence probability, or critical point \( p_c \), which marks the onset of an order-disorder phase transition, can be determined by analyzing the system in the \( \lim_{k \to 1/2} \) (see Appendix~\ref{app:A}).  

For \( s = 1 \), this model simplifies to the standard \( q \)-voter model with independence, as described in Ref.~\cite{nyczka2012phase}. Previous studies have shown that the \( q \)-voter model undergoes a discontinuous phase transition for \( q > 5 \) and a continuous phase transition for \( q \leq 5 \). Consistent with this result, our analysis demonstrates that the model exhibits a discontinuous phase transition for \( q > 5 \) and a continuous phase transition for \( q \leq 5 \) for all values of \( s \neq 0 \). Although the parameter \( s \) plays a crucial role in inducing phase transitions, it does not alter the nature of the transition (whether it is continuous or discontinuous). Instead, \( s \) primarily shifts the system's critical point \( p_c \).

\begin{figure}[t]
    \centering
    \includegraphics[width = 0.85\linewidth]{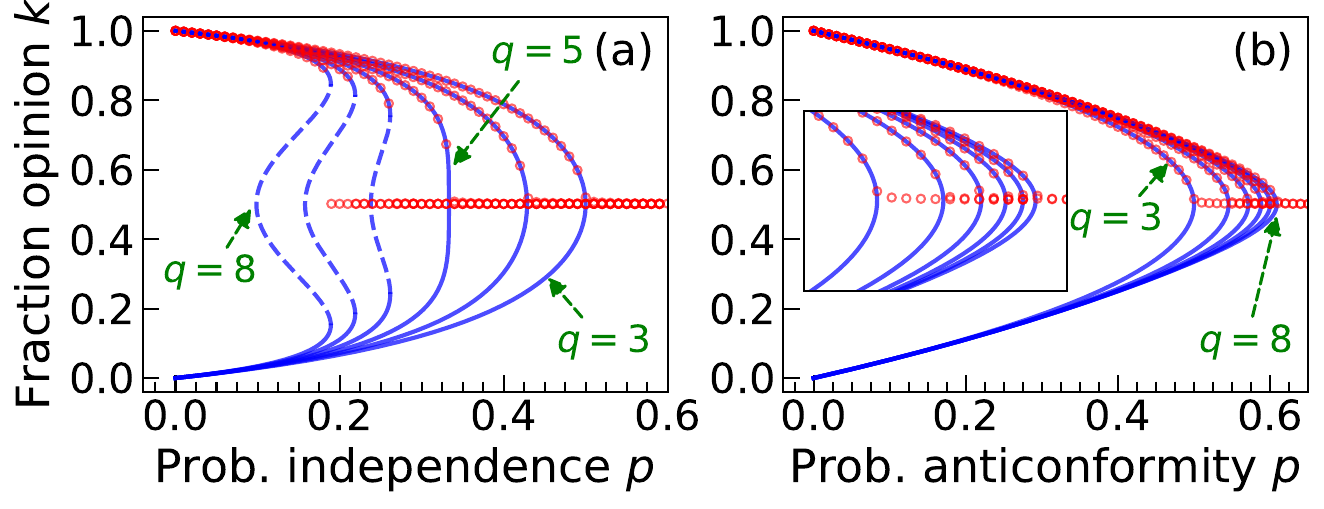}
    \caption{Phase diagram of the model with independence [panel (a), Eq.~\eqref{eq:Eq20}] shows a discontinuous phase transition for \( q > 5 \) (dashed lines) and a continuous transition for \( q \leq 5 \). In contrast, the model with anticonformity [panel (b), Eq.~\eqref{eq:Eq22}] exhibits a continuous phase transition for all \( q > 1 \). The inset graph provides a zoomed-in view for clarity. The red circle marker represents the MC simulation results, which align closely with the analytical results. The skepticism level is \( s = 1/2 \).}
    \label{fig:order_indep1}
\end{figure}

Figure~\ref{fig:order_indep1}(a) presents the plot of Eq.~\eqref{eq:Eq20} for various values of \( q \)-sized agents and a fixed skepticism level of \( s = 1/2 \). The results indicate that the model undergoes a continuous phase transition for \( q \leq 5 \) and a discontinuous phase transition for \( q > 5 \). The dashed line segment for \( q > 5 \) corresponds to the unstable region of the system.  In a social context, this phase diagram represents the transition between consensus and polarization phases. At \( p < p_c \), the system achieves consensus, reflecting agreement among agents or exhibiting a coexistence of majority and minority opinions. Complete consensus occurs in the absence of independence (\( p = 0 \)), as indicated by \( |m| = 1 \). As \( p \) increases, consensus decreases until the critical probability \( p_c \) is reached. At \( p \geq p_c \), the system is in a stalemate situation or polarization, where no single opinion dominates.

Similar to the model with independence, our analysis focuses on examining the impact of the probability \( p \) on the transition dynamics of skepticism \( s \). By analyzing the steady-state condition of Eq.~\eqref{eq:Eq18}, where \(\varphi_{+}(k) = \varphi_{-}(k)\), the probability of anticonformity \( p \) for a specific range of \( k \) can be expressed as:
\begin{align}\label{eq:Eq21}
   p  =  
  \begin{dcases} 
    \dfrac{k_{st}^q\left(1-k_{st}\right)-k_{st}\left(1-k_{st}\right)^q}{k_{st}^q-\left(1-k_{st}\right)^q\left[s\left(1-k_{st}\right)+k_{st}\right]} & \text{for} \, k \leq \dfrac{1}{2}, \\
  \dfrac{k_{st}\left(1-k_{st}\right)^q-k_{st}^q\left(1-k_{st}\right)}{k_{st}\left(1-k_{st}\right)^q+\left(1-k_{st}\right) \left(1-k_{st}^q\right)-s/2} & \text{for} \, k > \dfrac{1}{2}.
  \end{dcases}
\end{align}

\begin{figure}[tb]
    \centering
    \includegraphics[width = 0.95\linewidth]{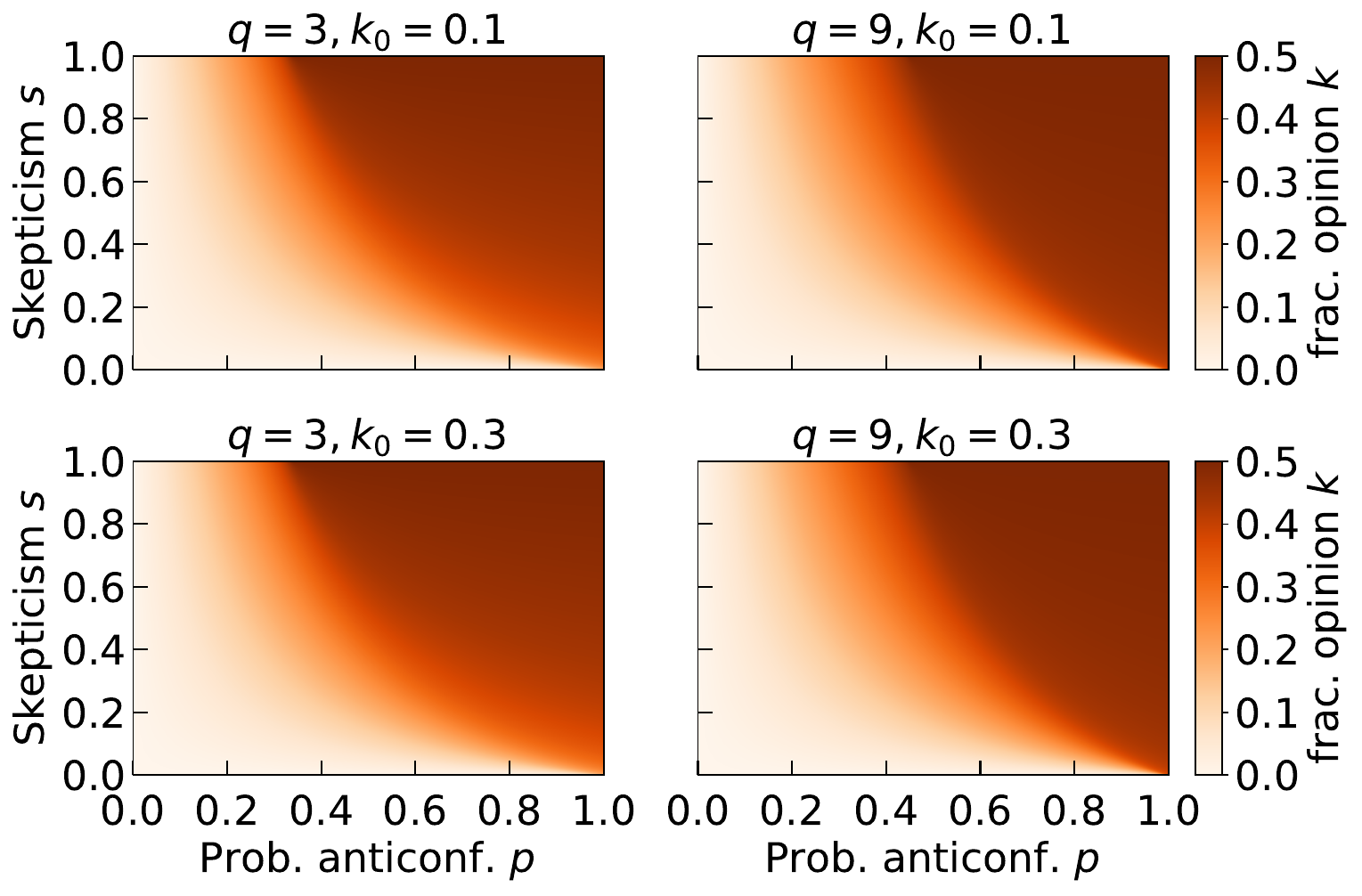}
    \caption{ The effect of the skepticism $s$ and the anticonformity $p$  to the outcome of the fraction opinion $k$ based on Eq.~\eqref{eq:Eq21} for $k \leq 1/2$. One can see that the fraction opinion $k$ increases as $s$  and $p$ increase. For both $q =  3$ and $q = 9$, the transition of the skepticism $s$  is smooth at all ranges $p$.}
    \label{fig:order_map}
\end{figure}

Based on Eq.~\eqref{eq:Eq21}, the stationary concentration \( k \) is influenced by the probability \( p \), the skepticism level \( s \), and the size of the \( q \)-sized agent group. Similar to the case where \( k \leq 1/2 \), our analysis shows that the transition of skepticism \( s \) is smooth for both \( q = 3 \) and \( q = 9 \) across all ranges of \( p \), regardless of whether \( k_0 = 0.1 \) or \( k_0 = 0.3 \) (see Fig.~\ref{fig:order_map}).   This finding contrasts with the independence model discussed in Ref.~\cite{civitarese2021external}, where the transition of \( s \) is non-smooth for larger \( q \). The smooth transition observed here suggests the presence of a continuous phase transition in this model. Furthermore, the impact of probability \( p \) on skepticism \( s \) differs from the model with independence, where a non-smooth (discontinuous) transition of \( s \) is observed for \( q = 9 \) at low \( p \) values~\cite{civitarese2021external}.  

Similarly to the model with independence, by applying the stationary condition of Eq.~\eqref{eq:Eq18}, the anticonformity probability \( p \) at the equilibrium state can be expressed as:
\begin{equation} \label{eq:Eq22}
p = \left[s\left(\dfrac{k_{st}^{q+1}-\left(1-k_{st}\right)^{q+1}}{k_{st}^q\left(1-k_{st}\right)-k_{st}\left(1-k_{st}\right)^q}\right)+1\right]^{-1}.
\end{equation}
The plot of Eq.~\eqref{eq:Eq22} for various values of \( q \)-sized agents with \( s = 1/2 \) is presented in Fig.~\ref{fig:order_indep1}(b). The results show that the model undergoes a continuous phase transition for all values of \( q \)-sized agents. As described in Eq.~\eqref{eq:Eq22}, the stationary probability $p$ depends not only on \( q \) but also on the skepticism level \( s \), decreasing as \( s \) increases. For the special case where \( s = 1 \), the model simplifies to the one discussed in Ref.~\cite{nyczka2012phase}.

In general, Eqs.~\eqref{eq:Eq20} and \eqref{eq:Eq22} can be simplified and expressed in the form \( m \sim (p - p_c)^{\beta} \), after re-scaling \( k \to (m + 1)/2 \), where \( p_c \) is the critical point of the model and \(\beta = 1/2\) holds for all values of \( q \neq 1 \).  For example, in the model with independence and \( q = 2 \), the magnetization \( m(p, s) \) is given by $m(p, s) = \pm \left[(1 - \left(1 + 2s\right)p)/(1 - p)\right]^{1/2}$, where the critical point is \( p_c = 1/(1 + 2s) \), corresponding to the condition \( m = 0 \). The critical exponent \(\beta = 1/2\) matches that of the mean-field Ising model, leading to a collapse of all data near the critical point \( p_c \) \cite{gitterman2013phase}. Similarly, for the model with anticonformity, the same critical exponent \(\beta = 1/2\) is observed based on Eq.~\eqref{eq:Eq22}. Other critical exponents, such as those related to susceptibility (\(\chi\)) and the Binder cumulant (\( U \)), will be determined using MC simulations in the next section.

\begin{figure}[tb]
\centering
\includegraphics[width=\linewidth]{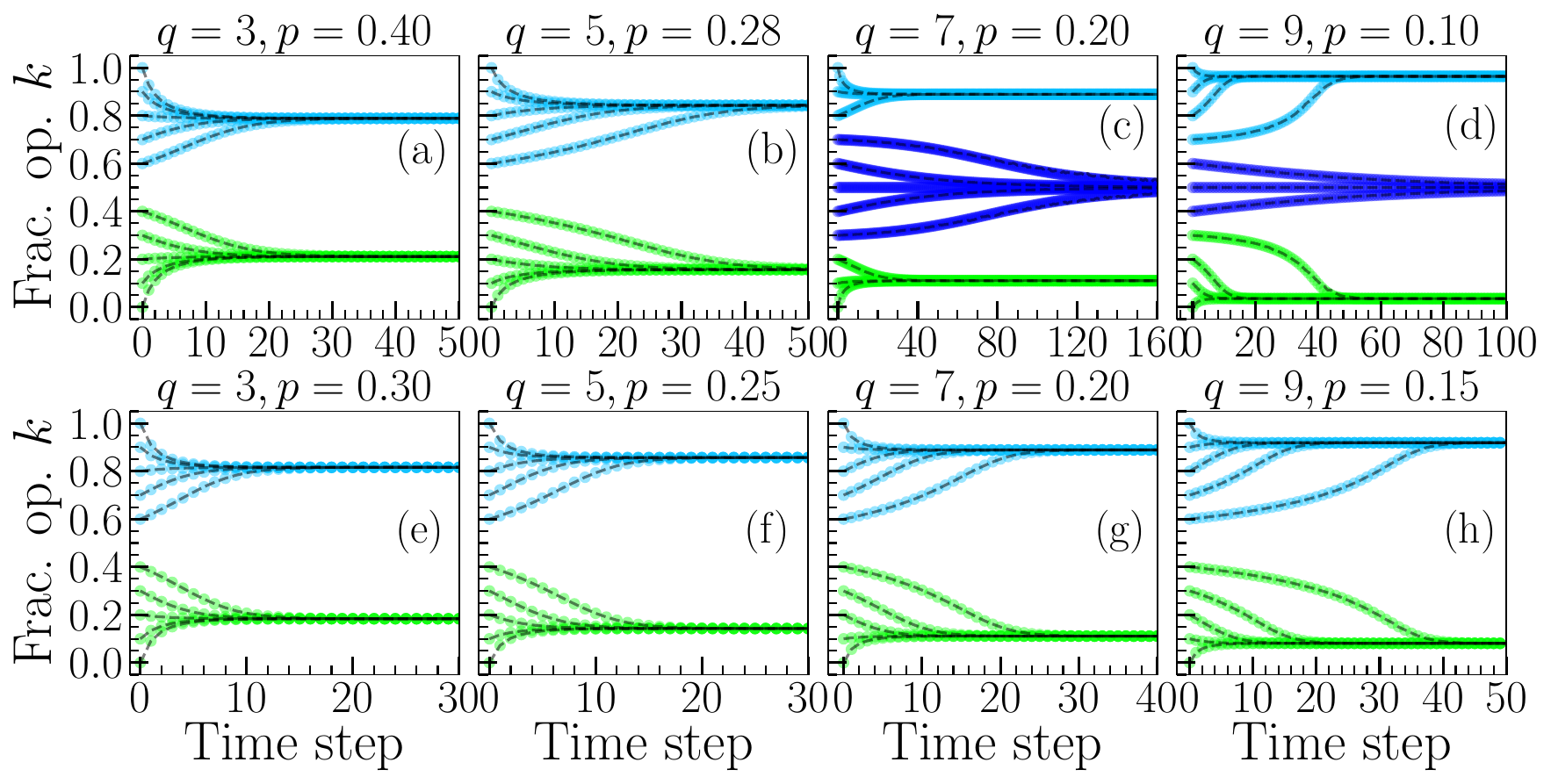}
\caption{\label{fig:tim_evol_indep} Time evolution of the opinion fraction \( k \) for the model with independence [(a)–(d)] and anticonformity [(e)–(h)] for different parameters: \( q = 3, 5, 7 \), and \( q = 9 \). In the model with independence, three stable states emerge for \( q = 7 \) and \( q = 9 \), indicating a discontinuous phase transition. In contrast, the model with anticonformity exhibits only bistable states for all values of \( q \), indicating a continuous phase transition. Data points represent MC simulations, while dot-dashed lines represent the RK solution of Eq.~\eqref{eq:Eq18}, with skepticism \( s = 1/2 \).}
\end{figure}

The time evolution of the fraction of opinions \( k \) can be analyzed using Eq.~\eqref{eq:Eq18}. While obtaining an exact analytical solution of Eq.~\eqref{eq:Eq18} for \( k \) as a function of time \( t \), i.e., solving \(\int \mathrm{d}k/[\varphi_{+}(k)-\varphi_{-}(k)] = \int \mathrm{d}t\), for arbitrary values of \( q \)-sized agents and \( s \), is challenging, a numerical approach provides a more practical alternative. Specifically, Eq.~\eqref{eq:Eq17} can be solved numerically using efficient techniques, such as the Runge-Kutta 4th order (RK) method \cite{pinder2018numerical}.  Figure~\ref{fig:tim_evol_indep} compares the MC simulation results (data points) with the RK method (dashed lines) for the models with independence and anticonformity, considering typical values of \( q \) and \( p < p_c \). The comparison reveals strong agreement between the MC simulation and the RK numerical solutions.  For \( q = 3 \) and \( q = 5 \), the fraction of opinions \( k \) evolves to two stable states, indicating a continuous phase transition. Conversely, for \( q = 7 \) and \( q = 9 \), \( k \) evolves to three stable states, reflecting a discontinuous phase transition. These results corroborate the order parameter transitions discussed in the previous section [see Fig.~\ref{fig:order_indep1}(a)].

The bottom panel of Fig.~\ref{fig:tim_evol_indep} illustrates the time evolution of the fraction of opinions \( k \) for the model with anticonformity. It is evident that \( k \) evolves to two stable states for all values of \( q \), indicating a continuous phase transition across all \( q \), as discussed in the previous section [see Fig.~\ref{fig:order_indep1}(b)]. The stability of these states can be further analyzed using the system's effective potential and the corresponding probability density function, which will be explored in detail in the next section.

\subsection{\label{sec:scaling_b} Scaling behavior}

To analyze the scaling behavior of the model, we formulate a scaling parameter to ensure data collapse across various values of \( s \). The procedure follows the method introduced in Ref.~\cite{sznajd2011phase}, which determines the scaling parameter for macroscopic properties of the model, including the order parameter \( m \), susceptibility \( \chi \), and Binder cumulant \( U \).  

The analysis begins by initializing the system in a disordered state, where the number of opinions \( +1 \) and \( -1 \) are equal. A group of \( q \)-sized agents and another individual agent (the voter) are then randomly selected to interact according to the model's dynamics. The voter follows three interaction rules, as outlined below:  
\begin{enumerate}
    \item[(1)] With probability \( p \), the voter acts as a nonconformist (either as an anticonformist or independent), and with probability \( s \), the voter acts skeptically.
    \item[(2)] With probability \( 1 - p \), the voter acts as a conformist, adopting the \( q \)-sized agent's opinion.
    \item[(3)] With probabilities \( p(1 - s) \) and \( p(1 - s/2) \), the voter does not change their opinion, respectively.
\end{enumerate}
Rule (3) does not change the system's state. Consequently, the scaling parameter primarily depends on the probability of nonconformity \( p \) and skepticism \( s \).  

During the dynamic process, the voter follows one of the three rules described above, generating a sequence of states where each state is determined by the initial conditions and the probabilities associated with rules (1) and (2), which are specific to the model. We can assert that if two systems have different parameters—say, system one is described by \((p_1, s_1)\) and system two by \((p_2, s_2)\)—but both systems share the same probability ratio \( r_1 = r_2 \), where \( r \) is defined as the ratio of the probability of rule (1) to rule (2). Both systems will statistically produce similar sequences of events, provided that their initial conditions are identical. 

To compare the two systems, we need to obtain the average time for the voter to change their opinion as a function of the parameters \( p \) and \( s \). We must calculate the probability of a change occurring in each iteration to determine the average time. The probability of a change occurring in the first iteration is \( p' = ps/2 + (1 - p) \) for the model with independence and  \( p' = ps + (1 - p) \) for the model with anticonformity. If no change occurs in the first iteration, with probability \( 1 - p' \), then in the second iteration, a change will occur with probability \( 2p'(1 - p') \). If no change occurs in the first and second iterations, the probability of a change occurring in the third iteration is \( 3p'(1 - p')^2 \), and so on. Thus, the average time can be expressed as an infinite series of all possible times required for a change to occur, which can be written as:
\begin{align}\label{eq:average_time}
    \tau = p'\sum_{j = 1}^{\infty} j \left(1-p'\right)^{j-1} = p' \sum_{j = 1}^{\infty} j q^{1-j} = \dfrac{p'}{\left(1-q\right)^2} = \dfrac{1}{p'}
\end{align}
where \( q = 1 - p' \). Consequently, based on Eq.~\eqref{eq:average_time}, the system's average time \( \tau \) is inversely proportional to the probability that causes the voter to change their opinion. This relationship aligns with the discussion in Ref.~\cite{sznajd2011phase}. 

If two systems share the same probability ratio \( r \) and state at a given scaled time \( t/\tau \), then their macroscopic parameters will also be identical, namely \( m_1 = m_2 \), \( \chi_1 = \chi_2 \), and \( U_1 = U_2 \). Consequently, by plotting  
\begin{align}\label{eq:Eq23}
     \text{prob. rule (1)}/p' \quad \text{or} \quad \text{prob. rule (2)}/p' \quad \text{vs.} \quad m, \chi, U, 
\end{align}  
the data will collapse for all values of \( s \).  These scaling parameters apply to any system that statistically adheres to the above rules. This paper tests the model on the complete graph and the scale-free network, as discussed in the next section. 

\subsection{\label{subsec:pot} Landau potential of the phase transition}
The Landau theory of free energy provides a framework for analyzing equilibrium phase transitions in thermodynamic systems \cite{landau1937theory, plischke1994equilibrium}. In his hypothesis, Landau proposed that the free energy can be expanded as a power series near the critical point regarding the order parameter. This approach can also be extended to analyze nonequilibrium systems, such as those described by the Langevin equation for two absorbing states under the mean-field approximation \cite{al2005langevin, vazquez2008systems}.  

In general, the Landau potential is not restricted to thermodynamic parameters like temperature, pressure, and volume; it can also depend on the system's order parameters, as defined in Eq.~\eqref{eq:Eq3}. We utilize the Landau potential to analyze the order-disorder phase transition in our model. According to Landau theory, the potential \( V \) can be expressed as $V(m) = \sum_{i=0}^{n} V_i m^i$, where \( m \) is defined as in Eq.~\eqref{eq:Eq3}, and \( V_i \) typically depends on the macroscopic parameters of the system, such as the probability \( p \) and skepticism \( s \). Since the potential \( V(m) \) is symmetric under inversion \( m \to -m \), all odd terms in the expansion vanish.  

To analyze the order-disorder phase transition of the model, we consider at least the first two even terms of the potential \( V(m) \), expressed as:  
\begin{equation}\label{eq:Eq24}
    V(m) = V_1 m^2 + V_2 m^4,
\end{equation}  
where the explicit forms of \( V_1 \) and \( V_2 \) depend on the specific model. The conditions for maximum and minimum values, which describe the order-disorder phase transitions, can be derived from the extremum conditions of Eq.~\eqref{eq:Eq24}. Furthermore, the nature of the phase transition—whether continuous or discontinuous—can be analyzed by choosing appropriate values for \( V_1 \) and \( V_2 \). Continuous and discontinuous phase transitions are possible for \( V_1 = 0 \). At the critical point, a continuous phase transition occurs if \( V_2 \geq 0 \) and a discontinuous phase transition occurs if \( V_2 < 0 \). 

To derive the effective potential-like function \( V(k) \) of the system, we utilize the probability densities of the fraction opinion \( k \) increasing and decreasing, as given by Eqs.~\eqref{eq:Eq8} and \eqref{eq:Eq9} for the model with independence, and Eqs.~\eqref{eq:Eq14} and \eqref{eq:Eq15} for the model with anticonformity. These probabilities define a force-like component of the model. In general, the potential \( V(k) \) can be expressed as $V(k) = -\int f(k) \, \mathrm{d}k$, where \( f(k) = \varphi_+(k) - \varphi_-(k) \) represents the effective force that drives the voter to change their opinion oppositely \cite{nyczka2012opinion}.

\begin{figure}[tb]
    \centering
    \includegraphics[width=\linewidth]{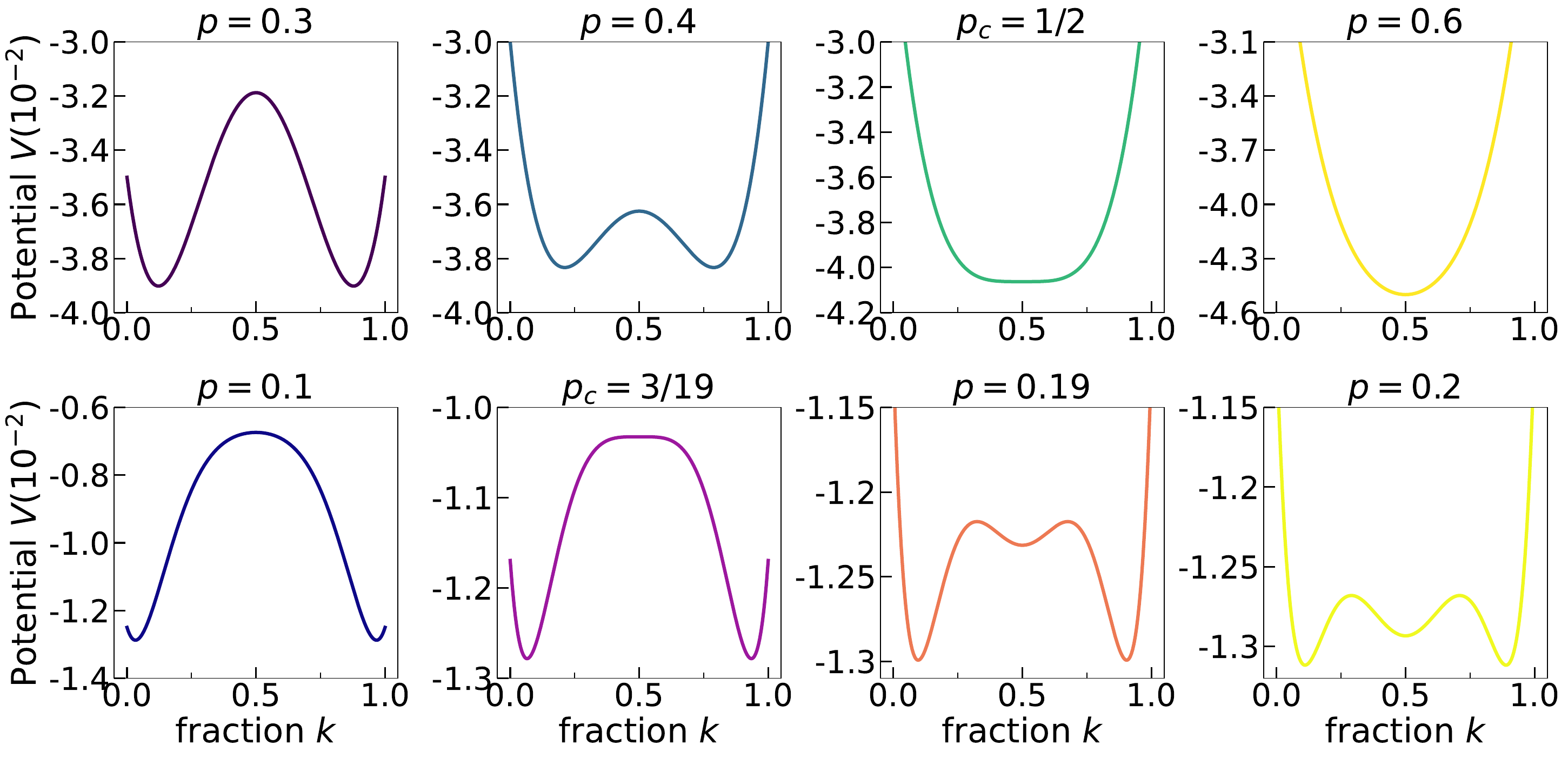}
    \caption{The potential \( V \) in Eq.~\eqref{eq:Eq25} illustrates the order-disorder phase transitions for the model with independence, considering typical values of \( q \) and \( s = 1/2 \). The top panels (\( q = 3 \)) show a continuous phase transition with the critical point at \( p_c = 1/2 \), where two minima are observed for \( p < p_c \), and one minimum is present for \( p > p_c \). The bottom panels (\( q = 7 \)) depict a discontinuous phase transition, with the critical point at \( p_c = 3/19 \). For \( p < p_c \), two minima exist, while for \( p > p_c \), three minima are observed.}
    \label{fig:pot_indep}
\end{figure}

The general solution of the potential $V$ for the model with independence is given by:
\begin{align}\label{eq:Eq25}
    V(k,q,p,s) = & \left(1-p\right)\Big[\dfrac{k^{q+2}}{q+2}-\dfrac{k^{q+1}}{q+1}- \dfrac{\left(k\,q+k+1\right)}{\left(q+1\right)\left(q+2\right)}\nonumber \\
    & \times \left(1-k\right)^{q+1}\Big] -k\left(1-k\right)\dfrac{p\,s}{2}.
\end{align}
The plot of Eq.~\eqref{eq:Eq25} for \( q = 3 \) (top panels) and \( q = 7 \) (bottom panels), considering typical values of probability \( p \) and skepticism \( s = 1/2 \), is shown in Fig.~\ref{fig:pot_indep}.  For \( q = 3 \), the potential is bistable at \( k \neq 1/2 \) and unstable at \( k = 1/2 \) for \( p < 1/2 \), while it becomes monostable at \( k = 1/2 \) for \( p > 1/2 \). The transition from bistable to monostable occurs at \( k = 1/2 \) when \( p = 1/2 \). This potential behavior indicates that the system undergoes a continuous phase transition, with the critical point at \( p = p_c = 1/2 \). For \( q = 7 \), the potential is bistable at \( k \neq 1/2 \) and unstable at \( k = 1/2 \) for \( p < 3/19 \), while it exhibits three stable states at \( 0 < k \leq 1/2 \) for \( p > 3/19 \). The transition from bistable to three stable states occurs at \( k = 1/2 \) when \( p = 3/19 \). This potential behavior indicates that the model undergoes a discontinuous phase transition, with the critical point at \( p = p_c = 3/19 \). These potentials corroborate the time evolution of the fraction of opinion \( k \) shown in Fig.~\ref{fig:tim_evol_indep}. The fraction \( k \) evolves to two stable states for \( q = 3 \) and to three stable states for \( q = 7 \).

\begin{figure}[tb]
    \centering
    \includegraphics[width=\linewidth]{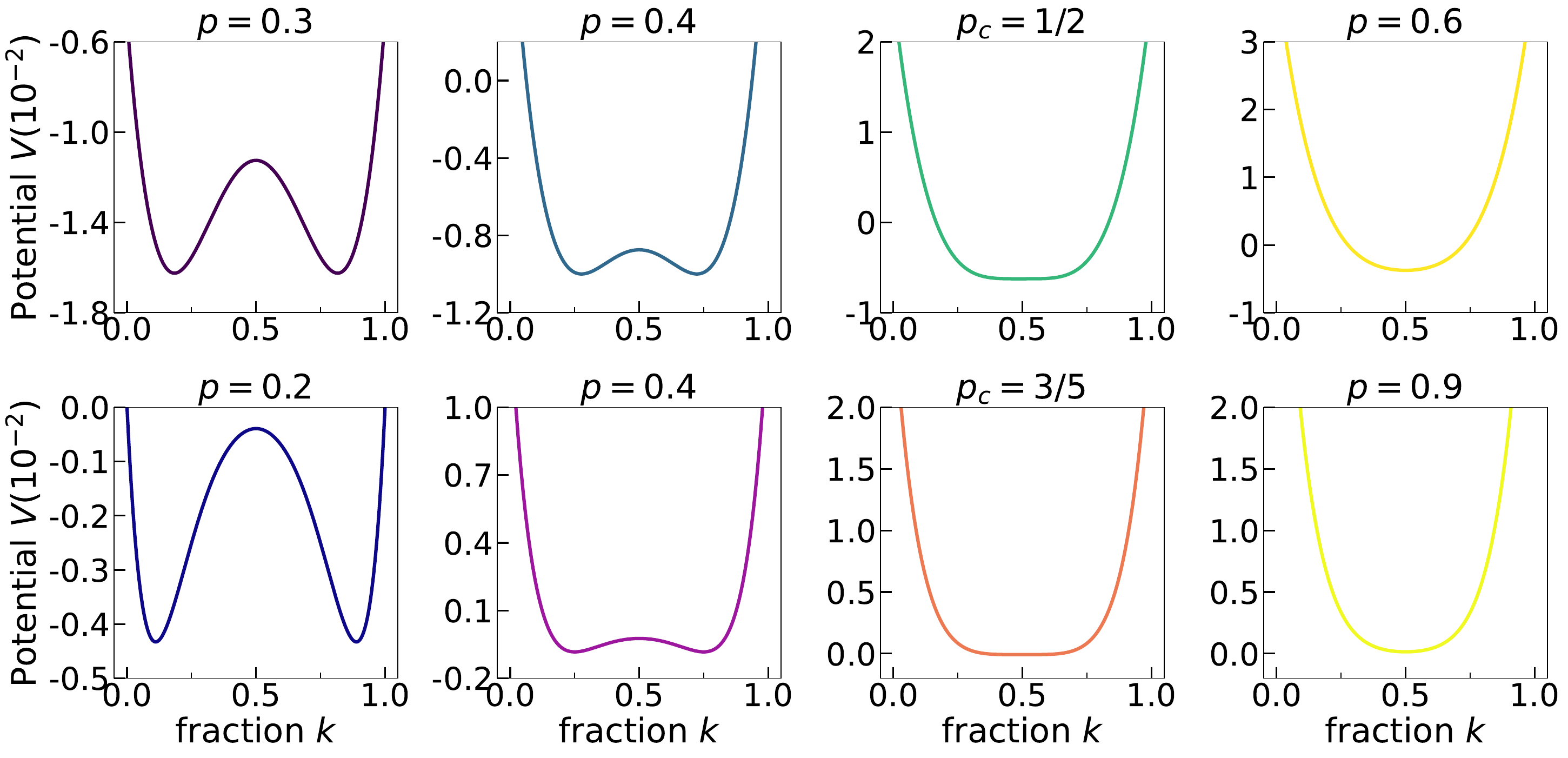}
    \caption{The potential \( V \) for the model with anticonformity, as described in Eq.~\eqref{eq:Eq26}, is shown for \( q = 3 \) (top panels), \( q = 7 \) (bottom panels), and \( s = 1/2 \). The top panels illustrate a continuous phase transition with the critical point at \( p_c = 1/2 \), characterized by two minima for \( p < p_c \) and one minimum for \( p > p_c \). Similarly, the bottom panels show a continuous phase transition with the critical point at \( p_c = 3/5 \), also exhibiting two minima for \( p < p_c \) and one minimum for \( p > p_c \).}
    \label{fig:pot_anti}
\end{figure}

For the model with anticonformity, the general solution for the potential \(V\) is given by:
\begin{align}\label{eq:Eq26}
     V(k,q,p,s) = & \dfrac{\left(1-p+ps\right)}{\left(q+2\right)}\left[k^{q+2}-\left(1-k\right)^{q+1} \right]-\dfrac{\left(1-p\right)}{\left(q+1 \right) \left( q+2 \right)} \nonumber \\ & \times \left[k^{q+1}-\left(1-k\right)^{q+1}\right] +\dfrac{ps\left(1-k\right)}{\left(q+2\right)}.
\end{align}
The plot of Eq.~\eqref{eq:Eq26} for \( q = 3 \) (top panels) and \( q = 7 \) (bottom panels), with typical values of \( p \) and \( s = 1/2 \), is shown in Fig.~\ref{fig:pot_anti}.  For both \( q = 3 \) and \( q = 7 \), the potential \( V \) exhibits two stable states at \( k \neq 1/2 \) for \( p < p_c \), transitioning to a monostable state for \( p > p_c \). This behavior indicates that the model undergoes a continuous phase transition for both values of \( q \). The critical points are \( p_c = 1/2 \) for \( q = 3 \) and \( p_c = 3/5 \) for \( q = 7 \).  This potential behavior aligns with the time evolution of the fraction of opinion \( k \) observed in the previous section (see Fig.~\ref{fig:tim_evol_indep}).

The potential in Eqs.~\eqref{eq:Eq25} and \eqref{eq:Eq26} reaches a maximum at \( k = 1/2 \) when \(\partial^2 V(k, q, p, s)/\partial k^2 \big|_{k = 1/2} < 0\), indicating the presence of two bistable states. Conversely, the potential reaches a minimum at \( k = 1/2 \) when \(\partial^2 V(k, q, p, s)/\partial k^2 \big|_{k = 1/2} > 0\), corresponding to a single stable state. The transition between these states occurs at the critical point, defined by the condition \(\partial^2 V(k, q, p, s)/\partial k^2 \big|_{k = 1/2} = 0\). This allows us to determine the system's critical point based on the model's parameters.  

For the model with independence, the critical point is given by: 
\begin{equation}\label{eq:pc_indep}
    p_c = \dfrac{\left(q - 1\right)2^{1-q}}{\left(q - 1\right)2^{1-q} + s}.
\end{equation}
Moreover, for the model with anticonformity, the critical point is given by:
\begin{equation}\label{eq:pc_anti}
    p_c = \dfrac{q - 1}{q - 1 + \left(q+1\right)s}.
\end{equation}

The potential \( V \) in Eqs.~\eqref{eq:Eq25} and \eqref{eq:Eq26} is expressed as a function of the fraction opinion \( k \). However, it can be transformed into a function of the order parameter \( m \) by substituting \( k = (m + 1)/2 \) into Eqs.~\eqref{eq:Eq25} and \eqref{eq:Eq26}, resulting in a form analogous to the Landau potential in Eq.~\eqref{eq:Eq24}.  For the model with independence, we derive the following expression for \( V_1 \), namely $V_1(q, p, s) = ps/4 - [(q - 1)(1 - p)]/2^{1 + q}$. Setting \( V_1(q, p, s) = 0 \), we obtain the critical probability as $p_c = (q - 1)2^{1 - q}/[(q - 1)2^{1 - q} + s] $,  which is consistent with Eq.~\eqref{eq:pc_indep}.  Additionally, at \( p = p_c \), the coefficient \( V_2 \) becomes: 
\begin{equation}\label{eq:V2_indep}
 V_2(q,s) =  -\frac{q\left(q -1\right) \left(q -5\right) s}{4\left(q-1\right)+2^{q+1} s },   
\end{equation}
which is always positive for \(1 < q \leq 5\), indicating a continuous phase transition and negative for \(q > 5\), indicating a discontinuous phase transition for all values of \(s \neq 0\).

For the model with anticonformity, the coefficient \( V_1(q, p, s) \) is given by $ V_1(q, p, s) = \left[\left(q + 1\right)ps - \left(q - 1\right)\left(1 - p\right)\right]2^{-(q+1)}$. Setting \( V_1(q, p, s) = 0 \), we obtain the critical point as $ p_c = (q - 1)/[q - 1 + \left(q + 1\right)s] $, which is consistent with \( p_c \) in Eq.~\eqref{eq:pc_anti}.  At \( p = p_c \), the term \( V_2(q, s) \) is always positive for all \( q \geq 2 \) and \( s \neq 0 \), given by:  
\begin{align}\label{eq:V2_anti}
    V_2(q,s) = \dfrac{qs \left(q^2-1\right) 2^{1-q}}{q-1+\left(q+1\right) s} > 0,
\end{align}
indicating a typical continuous phase transition for all values of $q\geq 2, s \neq 0$.

\begin{figure}[tb]
    \centering
    \includegraphics[width=0.55\linewidth]{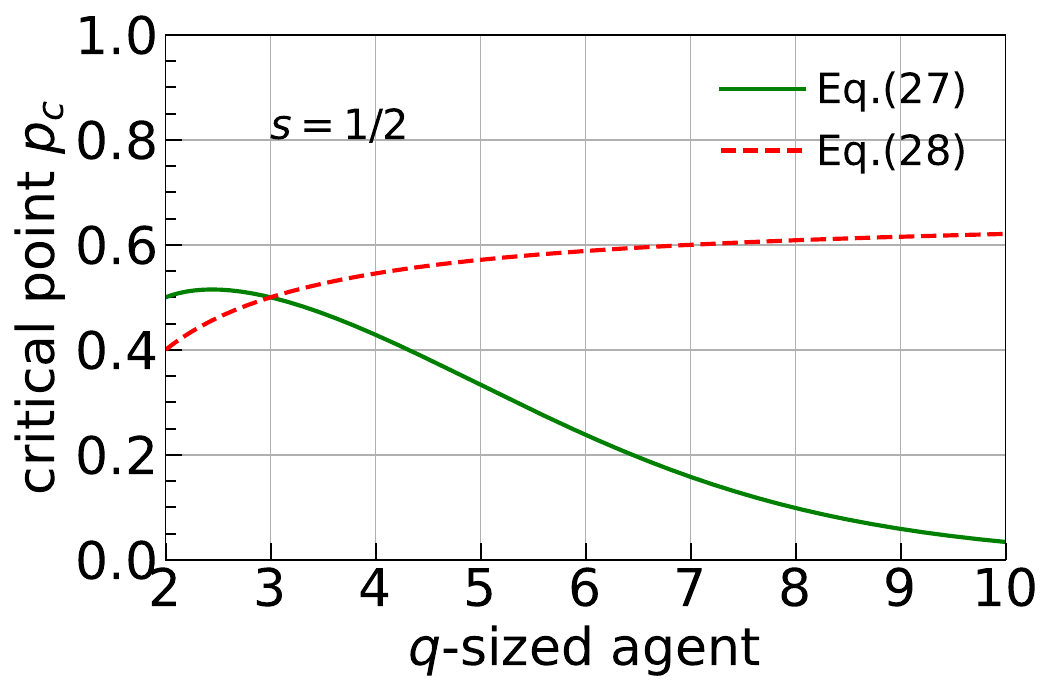}
    \caption{The plot of Eq.~\eqref{eq:pc_indep} and Eq.~\eqref{eq:pc_anti} for typical values of skepticism \( s = 1/2 \). The critical point of the model with independence is higher than that of the model with anticonformity for \( q < 3 \), and they converge at \( q = 3 \). For \( q > 3 \), the critical point of the model with independence is consistently lower than that of the anticonformity model.}
    \label{fig:pc_vs_q} 
\end{figure}

In general, the critical point of the model in Eqs.~\eqref{eq:pc_indep} and \eqref{eq:pc_anti} can be expressed in a simplified form as:  
\begin{equation}\label{eq:critical_var_s}
    p_c = \left(1 + c\,s\right)^{-1},
\end{equation}  
where \( c \) is a constant dependent on the \( q \)-sized agent.  For the model with independence, \( c = 2^{q-1}/(q-1) \) and for the model with anticonformity, \( c = (q+1)/(q-1) \).  From Eq.~\eqref{eq:critical_var_s}, we can observe the trend of the system's critical point concerning the parameter \( s \), as illustrated in Fig.~\ref{fig:pc_vs_q}. As shown in Fig.~\ref{fig:pc_vs_q}(a), for the independence model, the critical point \( p_c \) decreases as \( q \) increases. In contrast, for the anticonformity model [panel (b)], the critical point \( p_c \) increases with an increment in \( q \). Both models converge when \( s = 0 \), indicating the absence of an order-disorder phase transition.

\subsection{Fokker-Planck description of the phase transition}

In this section, we utilize the one-dimensional Fokker-Planck (FP) equation to analyze the order-disorder phase transition of the model, expressed as:  
\begin{equation}\label{eq:Eq27}
    \dfrac{\partial P(k,t)}{\partial t} = -\dfrac{\partial}{\partial k} \left[\xi_1(k)P(k,t) \right] + \dfrac{1}{2} \dfrac{\partial^2}{\partial k^2} \left[ \xi_2(k) P(k,t)\right],
\end{equation}  
where \(\xi_1 = (\varphi_+ - \varphi_-)\) and \(\xi_2 = (\varphi_+ + \varphi_-)/N\) represent the drift-like and diffusion-like coefficients of the model, respectively. 

We focus on analyzing the stationary condition of Eq.~\eqref{eq:Eq27}, which corresponds to the system's equilibrium state. The general solution for the stationary condition is given by:  
\begin{equation}\label{eq:Eq32}
 P_{\text{st}}(k) = \dfrac{C}{\xi_2} \exp\left[\int 2\dfrac{\xi_1}{\xi_2}\,\mathrm{d}k \right],
\end{equation}  
where \( C \) is a normalization constant ensuring that \( \int P_{\text{st}}(k)\, \mathrm{d}k = 1 \).  Finding an exact analytical solution for Eq.~\eqref{eq:Eq32} across different \( q \)-sized agents, probabilities \( p \), and skepticism values \( s \) is challenging. However, numerical methods can be employed to explore the stationary distribution \( P_{\text{st}}(k) \).

\begin{figure}[tb]
    \centering
    \includegraphics[width=\linewidth]{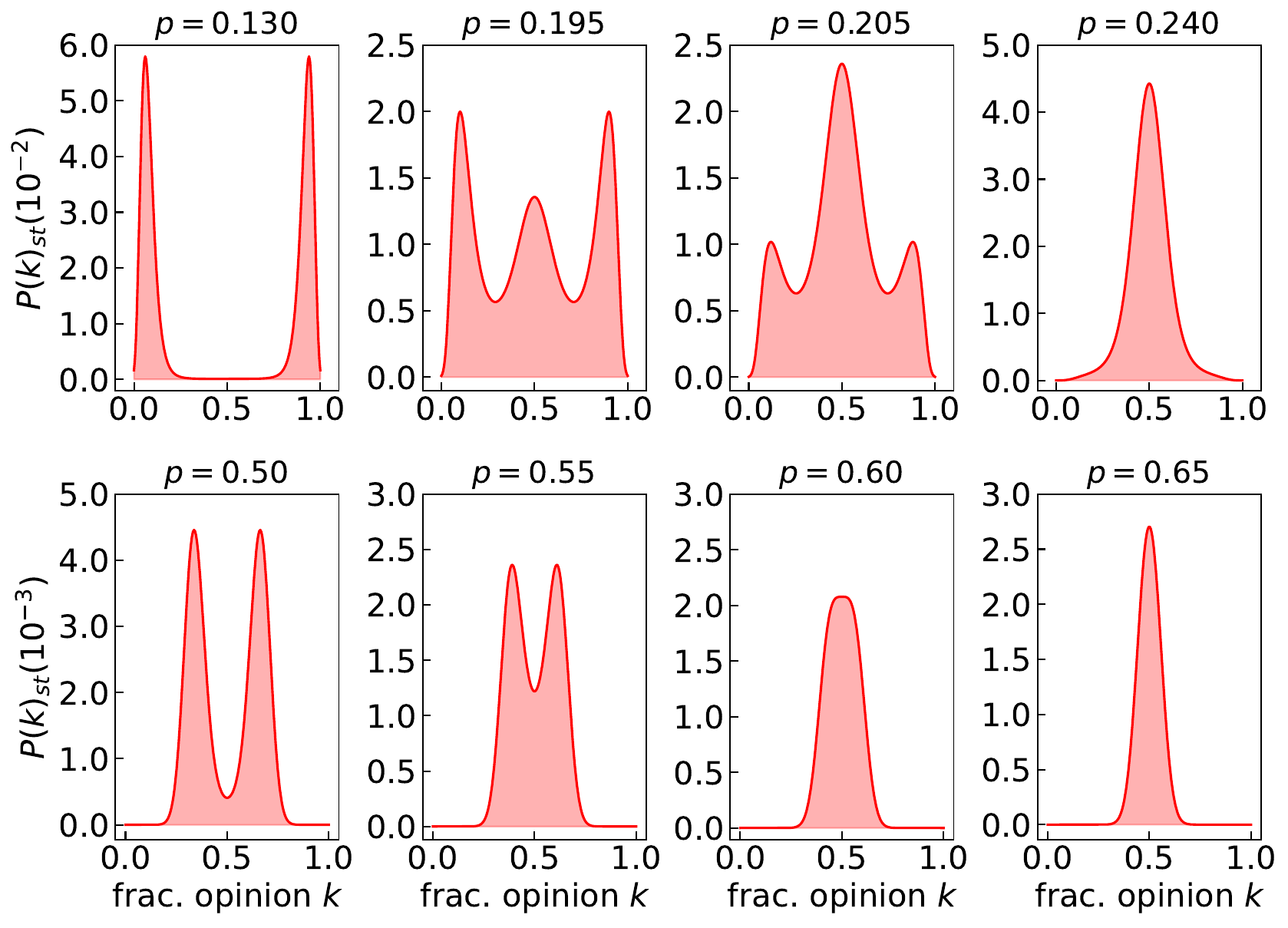}
    \caption{Stationary probability density function \( P_{\text{st}}(k) \) of the model with independence [top panels] based on Eq.~\eqref{eq:Eq32} for \( q = 7 \), \( s = 1/2 \), and typical values of \( p \). The function \( P_{\text{st}}(k) \) exhibits two peaks for \( p = 0.130 < p_c \), three peaks for \( p = 0.195 \) and \( p = 0.205 > p_c \), and one peak for \( p = 0.240 \). These peaks correspond to the minima of the effective potential described in Eq.~\eqref{eq:Eq25}, where the potential shows bistable states for \( p < p_c \) and three stable states for \( p > p_c \). [Bottom panels] The probability density function \( P_{\text{st}}(k) \) for the model with anticonformity shows two maxima at low \( p < p_c \), corresponding to two stable states in the effective potential described in Eq.~\eqref{eq:Eq26}. As \( p \) increases, the peaks of \( P_{\text{st}}(k) \) diminish, ultimately resulting in a single maximum for \( p > p_c \) at \( k = 1/2 \). This behavior of \( P_{\text{st}}(k) \) indicates the occurrence of a continuous phase transition.}
    \label{fig:prob_indep}
\end{figure}

Figure~\ref{fig:prob_indep} (top panels) presents the numerical solution for the model with independence, highlighting the characteristics of a discontinuous phase transition. Specifically, \( P_{\text{st}}(k) \) exhibits three peaks corresponding to the three stable states in the effective potential described by Eq.~\eqref{eq:Eq25}. For lower values of \( p \), \( P_{\text{st}}(k) \) has two peaks, with the peak at \( k = 1/2 \) becoming more pronounced as \( p \) increases. At higher values of \( p \), the probability density function \( P_{\text{st}}(k) \) transitions to a single peak at \( k = 1/2 \), indicating that the system stabilizes at \( k = 1/2 \). Similarly, the numerical solution of Eq.~\eqref{eq:Eq32} for the model with anticonformity is shown in the bottom panels of Fig.~\ref{fig:prob_indep}. For \( p < p_c \), \( P_{\text{st}}(k) \) displays two peaks, corresponding to two stable states in the effective potential described by Eq.~\eqref{eq:Eq26}. As \( p \) increases, the peak of \( P_{\text{st}}(k) \) decreases in prominence, ultimately resulting in a single peak at \( k = 1/2 \) for \( p > p_c \). This behavior of \( P_{\text{st}}(k) \) signifies the occurrence of a continuous phase transition. 

\subsection{\label{sec:numeric} Critical exponents and universality class}
\subsubsection{Model on the complete graph}

\begin{figure}[t]
    \centering
    \includegraphics[width=0.9\linewidth]{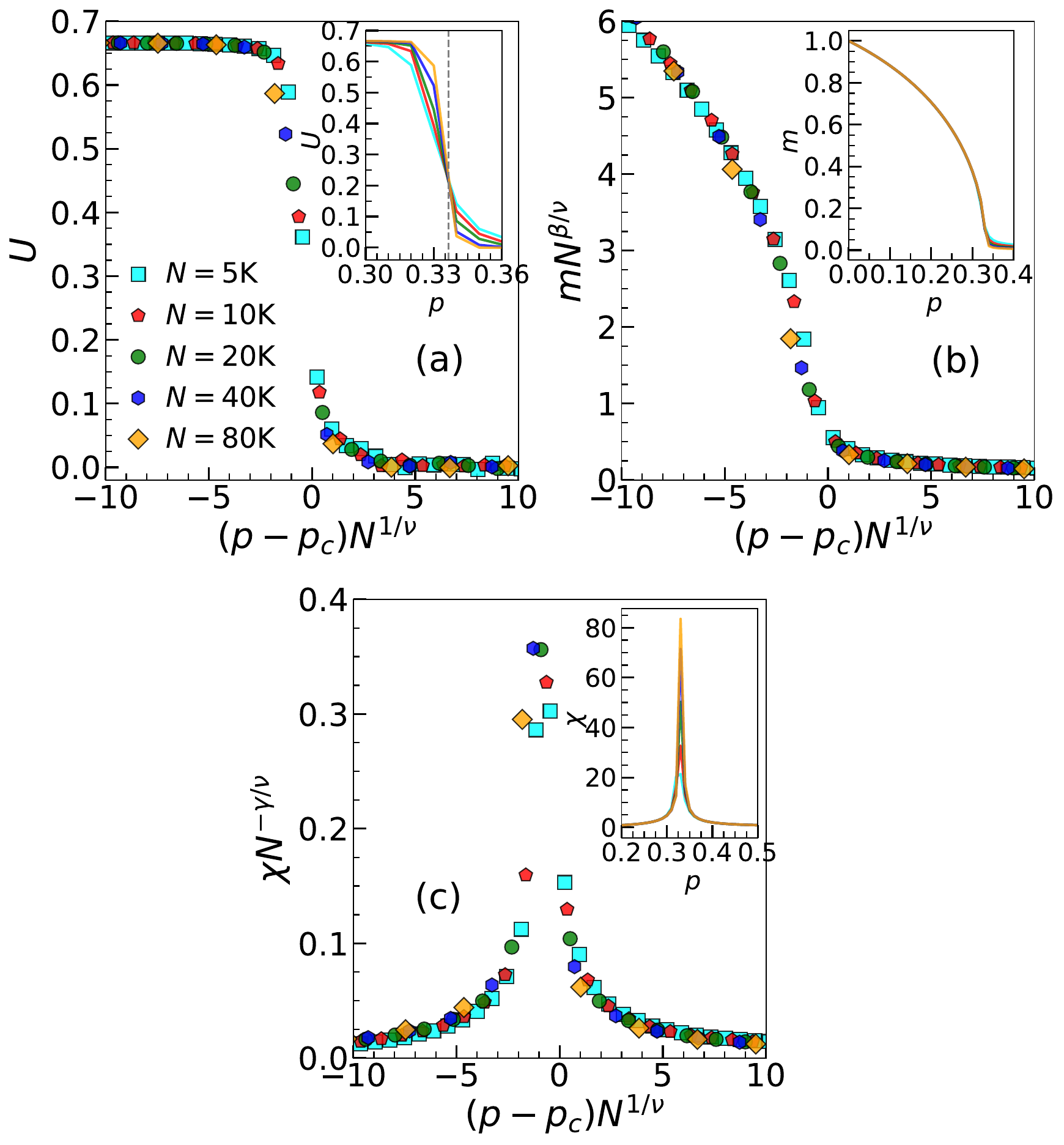}
    \caption{Numerical simulations of the model with independence showing the Binder cumulant \( U \) (a), order parameter \( m \) (b), and susceptibility \( \chi \) (c). The critical point \( p_c \approx 0.334 \) is determined from the intersection of lines in the \( U \) versus \( p \) plot [inset of panel (a)], indicating a continuous phase transition. Data collapse is observed for critical exponents \(\nu \approx 2.0\), \(\beta \approx 0.5\), and \(\gamma \approx 1.0\), confirming that the model belongs to the mean-field Ising universality class. The skepticism level is set to \( s = 1.0 \), and each data point represents an average of \( 10^5 \) independent realizations.}
    \label{fig:critical_mf_indep}
\end{figure}

We perform extensive numerical simulations to investigate the order parameter \( m \), Binder cumulant \( U \), and susceptibility \( \chi \) for the model with independence, focusing on a typical value of \( q = 3 \) and a skepticism level \( s = 1.0 \). The critical exponents \(\beta\), \(\nu\), and \(\gamma\) are determined using FSS analysis.  As shown in Fig.~\ref{fig:critical_mf_indep}, the model exhibits a continuous phase transition, with the critical point \( p_c \) determined from the intersection of lines in the Binder cumulant \( U \) versus probability \( p \) plot. The critical point is observed at \( p_c \approx 0.334 \), as illustrated in the inset of panel (a). This result aligns closely with the analytical prediction in Eq.~\eqref{eq:Eq20}, which gives \( p_c = 1/3 \) for \( q = 3 \) and \( s = 1 \).  

The main panels of Fig.~\ref{fig:critical_mf_indep} display the scaling plots near the critical point \( p_c \), showing data collapse for \(\beta \approx 0.5\), \(\nu \approx 2.0\), and \(\gamma \approx 1.0\). These critical exponents are consistent with those obtained for other opinion dynamics models, such as the Sznajd model  \cite{muslim2020phase, MUSLIM2022133379},  the majority rule model on the complete graph \cite{MUSLIM2022128307},  the kinetic exchange model \cite{crokidakis2014phase, krapivsky2010kinetic}, and the majority rule on complex networks \cite{mulya2024phase, muslim2024impact}. These critical exponents suggest that the model belongs to the same universality class as the mean-field Ising model. These scaling parameters are applicable in cases where the model undergoes a continuous phase transition (\( s \neq 0, 1 < q \leq 5 \)).

Our numerical simulation results for the model with anticonformity are consistent with the analytical results in Eq.~\eqref{eq:pc_anti} and the simulation results. For \( q = 4 \) and \( s = 0.5 \), the simulation yields a critical point of \( p_c \approx 0.374 \), which agrees with the analytical prediction of \( p_c = 3/8 \).  Using FSS analysis, we obtained the same critical exponents as those for the model with independence, namely \(\gamma \approx 1.0\), \(\beta \approx 0.5\), and \(\nu \approx 2.0\) (not shown). These critical exponents indicate that the model with anticonformist agents is identical to the model with independence in terms of scaling behavior and belongs to the same universality class as the mean-field Ising model.  

\subsubsection{The model on the scale-free network}

We performed extensive numerical simulations to study the model's phase transition and critical exponents. In this part of the study, we utilized the Barabási-Albert (BA) network, which is known for its degree distribution following a power law, \( P(\kappa) \sim \kappa^{-\zeta} \), where the exponent \( \zeta = 3 \) in the thermodynamic limit (\( N \to \infty \))~\cite{barabasi1999emergence, albert2002statistical}. This value of \( \zeta \) is specific to the BA model, a prototypical example of a scale-free network. In general, scale-free networks exhibit power-law exponents \( \zeta \) that can vary, often falling within the range \( 2 < \zeta < 3 \), depending on the network's generative mechanisms and parameters~\cite{albert2002statistical}. In the BA network, each node is configured to have at least a minimum number, \( q \), of nearest neighbors. A voter is then randomly selected from the network, along with \( q \) of its nearest neighbors, also chosen at random. The interactions between the voter and these \( q \) neighbors follow the model's algorithm.

\begin{figure}[tb]
    \centering
    \includegraphics[width = \linewidth]{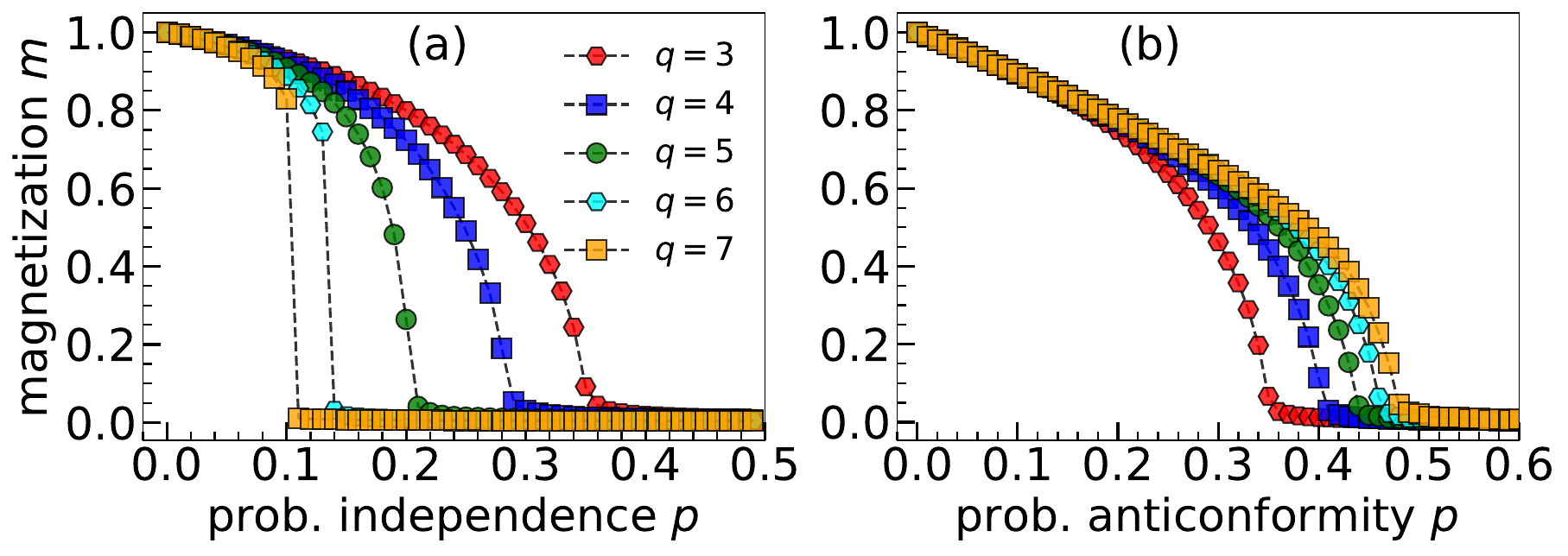}
    \caption{The phase diagrams for the model with independence [panel (a)] and for the model with anticonformity [panel (b)] on the BA network are presented for various values of \( q \) with \( s = 0.5 \). In the model with independence, a continuous phase transition is observed for \( q \leq 5 \), while a discontinuous transition occurs for \( q > 5 \). In contrast, the model with anticonformity  consistently displays a continuous phase transition across all \( q \) values. The average node degree of the network is $\langle \kappa \rangle = 2q$, with the degree distribution exponent \(\zeta\) are: $\zeta \approx 2.44$ for \( q = 3 \), $\zeta \approx 2.50$ for \( q = 4 \), $\zeta \approx 2.56$ for \( q = 5 , q = 6, q = 7\). The population size is \( N = 5 \times 10^4 \), with each data point averaged over \( 10^5 \) independent realizations.}
    \label{fig:var_q_scale-free} 
\end{figure}

The numerical results for magnetization \( m \) in two scenarios—the model with independence and the model with anticonformity—for typical values of \( q \) and \( s = 0.5 \) are shown in Fig.~\ref{fig:var_q_scale-free}. For the model with independence, a continuous phase transition is observed for \( 1 < q \leq 5 \), while a discontinuous phase transition occurs for \( q > 5 \). In contrast, the model with anticonformity exhibits a continuous phase transition for all values of \( q \). This behavior mirrors the results obtained when the model is applied to a complete graph.  For \( s = 1 \), the model discussed in this paper becomes identical to the model presented in Ref.~\cite{jkedrzejewski2017pair}. Our findings are further corroborated by the analytical results derived using the Pair Approximation method in Ref.~\cite{jkedrzejewski2017pair}, which demonstrate that the nature of the phase transition—whether continuous or discontinuous—in complex networks is solely determined by the nonlinear parameter \( q \). Specifically, a continuous phase transition is observed for \( 1 < q \leq 5 \), while a discontinuous phase transition occurs for \( q > 5 \).  As previously mentioned, the parameter \( s \) in this paper does not affect the type of phase transition but only shifts the critical point of the model. 

\begin{figure}[tb]
    \centering
    \includegraphics[width=\linewidth]{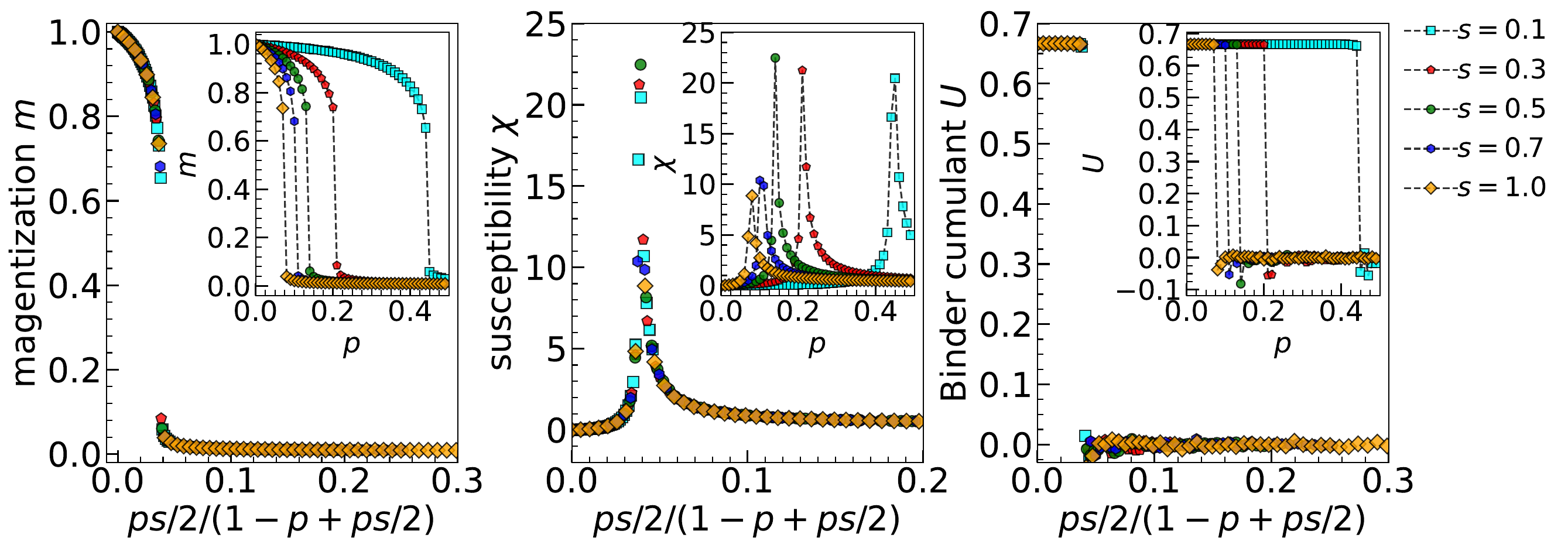}
    \caption{Scaling plot of the model with independence for \( q = 6 \) on the Barabási-Albert network with various values of skepticism \( s \). The model exhibits a discontinuous phase transition for all values of \( s \), with the critical point \( p_c \) increasing as \( s \) decreases (see inset graph). The main graph shows data collapse across all values of \( s \), confirming the robustness of the scaling relation. The average degree of the network is \( \langle \kappa \rangle = 12 \), with a degree distribution exponent \( \zeta \approx 2.49 \). The population size is \( 10^4 \), and each data point represents an average over \( 10^6 \) independent realizations.}
    \label{fig:indep_s_scale-free} 
\end{figure}

We investigated the scaling behavior of the independent voter model. The results from the numerical simulations for magnetization \( m \), susceptibility \( \chi \), and Binder cumulant \( U \) for \( q = 6 \) and several values of \( s \) are shown in Fig.~\ref{fig:indep_s_scale-free}. The inset graphs demonstrate that the model undergoes a discontinuous phase transition for all values of \( s \), with the critical point decreasing as \( s \) increases. This behavior aligns with the results from the complete graph model, where the parameter \( s \) only shifts the critical point \( p_c \). Furthermore, the main graphs illustrate the scaling plots for various values of \( s \). It can be observed that Eq.~\eqref{eq:Eq23} performs very well, as all data for \( s \) collapse for \( m \), \( \chi \), and \( U \).

\begin{figure}[t]
    \centering
    \includegraphics[width=0.9\linewidth]{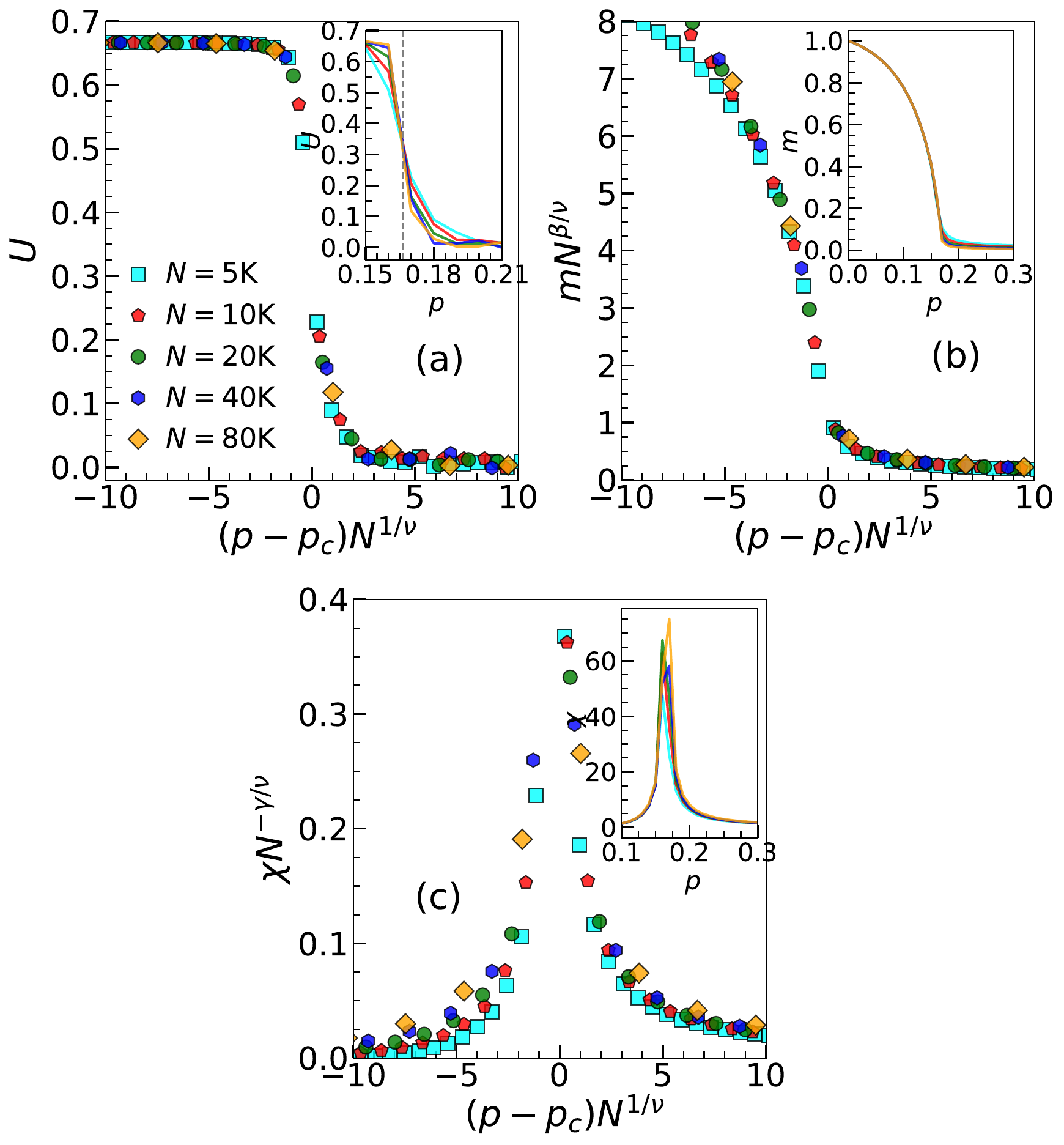}
    \caption{Numerical simulations of the model with independence on a scale-free network, showing the order parameter \( m \) (a), susceptibility \( \chi \) (b), and Binder cumulant \( U \) (c). The critical point \( p_c \approx 0.166 \) is determined from the intersection in panel (c), indicating a continuous phase transition. Data collapse is observed for critical exponents \( \nu \approx 2.0 \), \( \beta \approx 0.5 \), and \( \gamma \approx 1.0 \) (see insets), confirming that the model belongs to the mean-field Ising universality class. The average node degree of the network is \( \langle \kappa \rangle = 2q = 8 \), with skepticism level \( s = 1.0 \), and each data point represents an average over \( 10^6 \) independent realizations.}
    \label{fig:critical_exponents_indep_scare-free}
\end{figure}

To investigate the universality class of the model, we analyzed the model for \( q = 4 \) and \( s = 1.0 \) on a BA network with an average node degree of \( \langle \kappa \rangle = 8 \). The simulation results for the Binder cumulant \( U \), magnetization \( m \), and susceptibility \( \chi \) are shown in Fig.~\ref{fig:critical_exponents_indep_scare-free}.  From the Binder cumulant \( U \) versus the probability of independence \( p \), we observed a crossing of lines, indicating the critical point of the model at \( p_c \approx 0.166 \) [inset of panel (a)]. This critical point aligns well with the analytical prediction derived using the pair approximation method in Eq.~(34) of Ref.~\cite{jkedrzejewski2017pair}, which estimates \( p_c \approx 0.168 \). Using FSS analysis, the best-fitting critical exponents, which ensure data collapse for different system sizes \( N \) near the critical point \( p_c \), are determined as \( \beta \approx 0.5 \), \( \nu \approx 2.0 \), and \( \gamma \approx 1.0 \). These critical exponents are consistent with those reported for the \( q \)-voter model on random regular networks \cite{vieira2020pair}, suggesting that the model belongs to the same universality class as the mean-field Ising model.  Similar results have been obtained for the majority rule model for odd-sized groups of agents, defined on various complex networks such as Barabási-Albert, Erdős-Rényi, and Watts-Strogatz networks. These results also indicate that the model falls into the universality class of the mean-field Ising model \cite{muslim2024impact, mulya2024phase}.  

\begin{figure}[tb]
    \centering
    \includegraphics[width=\linewidth]{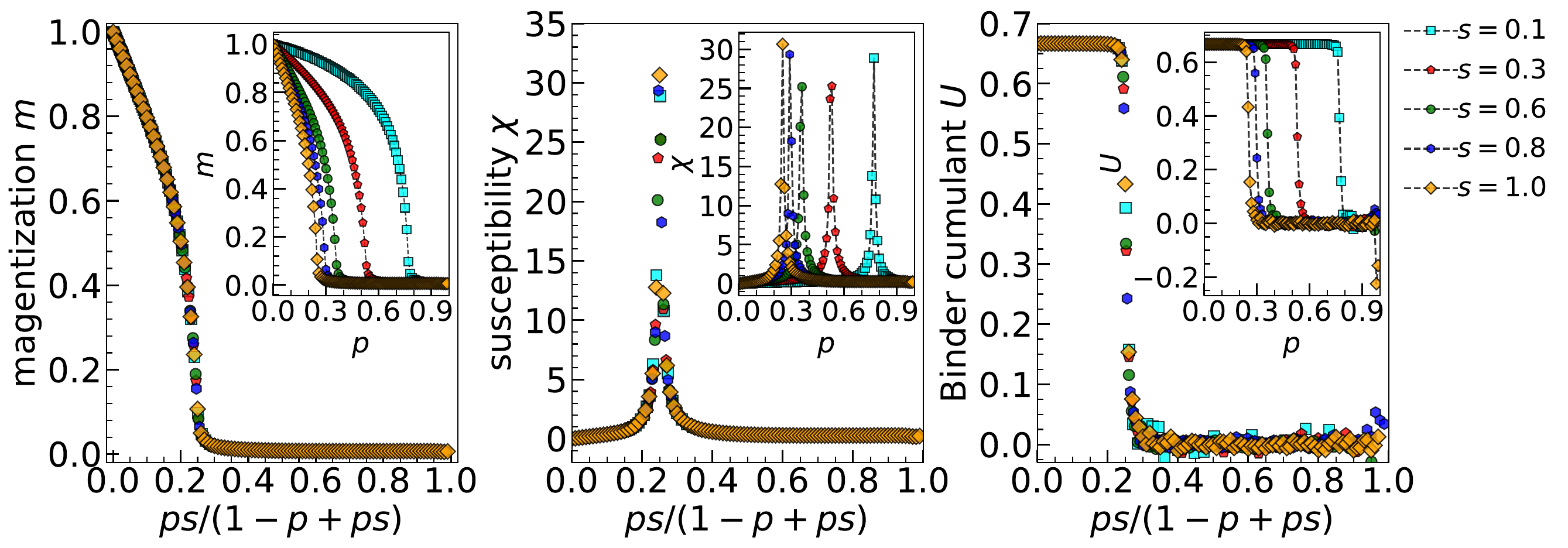}
    \caption{Scaling plot of the model with anticonformity for \( q = 4 \) on the BA network with various values of skepticism \( s \). The model exhibits a continuous phase transition for all values of \( s \), with the critical point \( p_c \) increasing as \( s \) decreases (see inset graph). The main graphs show excellent data collapse across all values of \( s \). The average degree of the network is \( \langle \kappa \rangle = 8 \), with a degree distribution exponent of \( \zeta \approx 2.44 \). The population size is \( 10^4 \), and each data point represents an average over \( 10^6 \) independent realizations.}
    \label{fig:anti_s_scale-free} 
\end{figure}

The numerical results for the model with anticonformity for several values of \( s \) are presented in Fig.~\ref{fig:anti_s_scale-free}. The model exhibits a continuous phase transition for all values of \( s \). The main graph displays the scaling plot based on Eq.~\eqref{eq:Eq23}, showing excellent data collapse across different values of \( s \). These findings indicate the model's scaling behavior is robust and consistent across various network structures, including 2D and 3D lattices (figures not shown). These results conclude that Eq.~\eqref{eq:Eq23} is independent of the underlying network structure. Instead, it depends solely on the parameters \( p \) and \( s \), or more generally, on the factors that govern changes in the voter's opinion during the dynamics. 

\begin{figure}[t]
    \centering
    \includegraphics[width=0.9\linewidth]{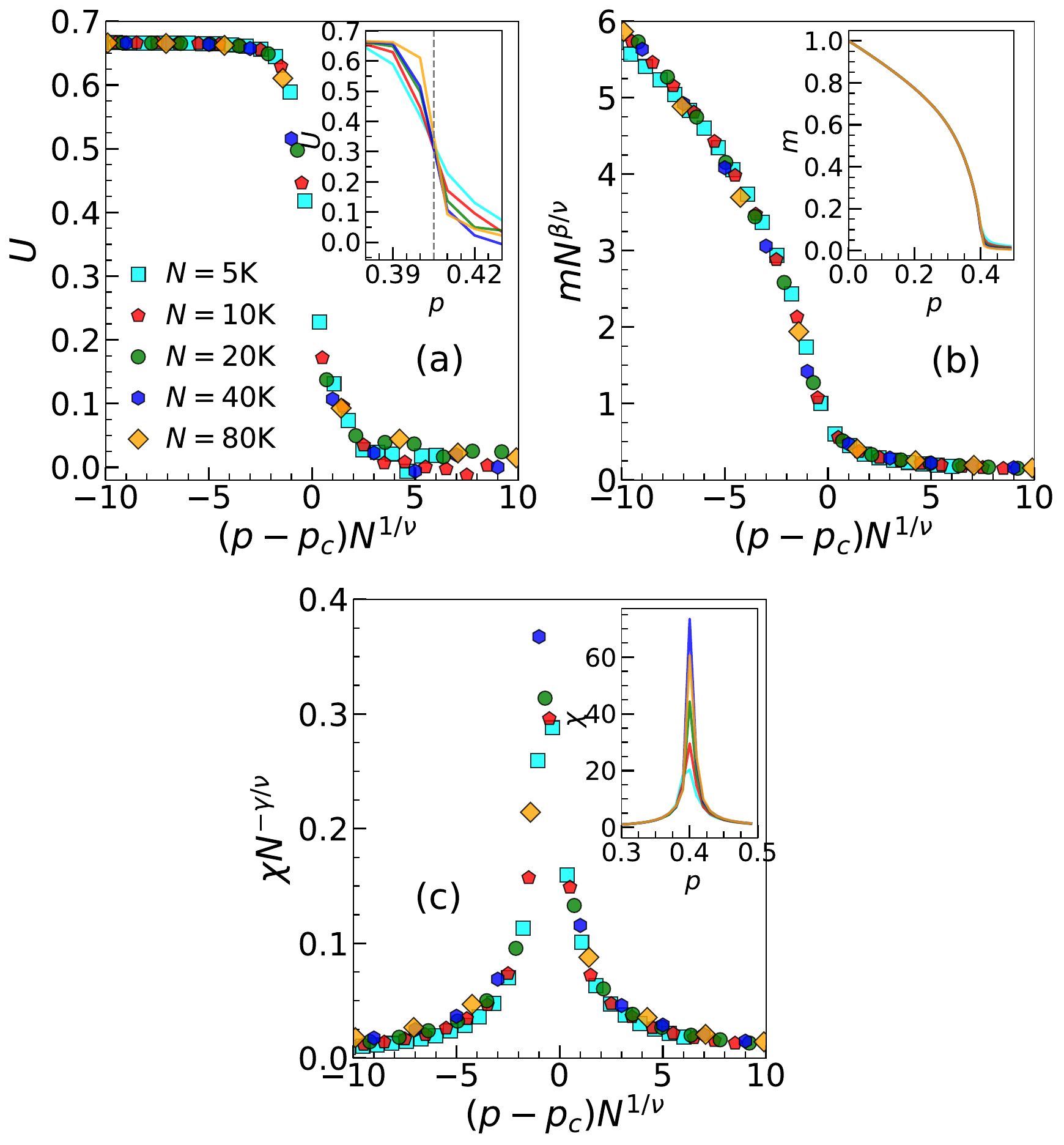}
    \caption{Numerical simulations of the model with anticonformity on the BA network, showing the order parameter \( m \) (a), susceptibility \( \chi \) (b), and Binder cumulant \( U \) (c). The critical point \( p_c \approx 0.405 \) is determined from the intersection in panel (c), indicating a continuous phase transition. Data collapse is observed for critical exponents \( \nu \approx 2.0 \), \( \beta \approx 0.5 \), and \( \gamma \approx 1.0 \) (see insets), confirming that the model aligns with the mean-field Ising universality class. The skepticism level is set to \( s = 0.5 \), and each data point represents an average over \( 10^6 \) independent realizations.}
    \label{fig:critical_exponents_anti_scare-free}
\end{figure}

We also examined the critical exponents of the model for \( q = 4 \) and \( s = 0.5 \), as shown in Fig.~\ref{fig:critical_exponents_anti_scare-free}. The inset graph of the panel (a) indicates that the critical point of the model occurs at \( p_c \approx 0.405 \), confirming that the model undergoes a continuous phase transition. The best-fitting critical exponents, which result in data collapse for different system sizes \( N \) near the critical point (as shown in the main panels), are determined as \( \beta \approx 0.5 \), \( \nu \approx 2.0 \), and \( \gamma \approx 1.0 \). These values indicate that the model belongs to the same universality class as the mean-field Ising model. Moreover, the critical exponents \( \beta \), \( \nu \), and \( \gamma \) satisfy the hyperscaling relation \( \nu' d_c = 2 \beta + \gamma \), where \( d_c = 4 \) is the critical dimension of the network. 

\section{\label{sec:Sec4} Final Remark}
%-------------------------------------------------------------------------------

This study investigated the nonlinear \( q \)-voter model on complete graphs and scale-free networks (Barabási-Albert network), focusing on two key social behaviors: anticonformity and independence. These behaviors were modeled using the parameter \( p \), representing the probability of a voter adopting either anticonformist or independent behavior. Additionally, a skepticism parameter \( s \) was introduced to capture the voter's skepticism toward the opinion of the \( q \)-sized agent.  We analyzed the impact of these social behaviors and skepticism on macroscopic phenomena, notably the order-disorder phase transition and the scaling behavior of the model. An analytical mean-field approach was employed to study the model on complete graphs, complemented by numerical simulations to explore the dynamics on both complete graphs and scale-free networks. 

The main findings of this study highlight the distinct behaviors of the nonlinear \( q \)-voter model on complete graphs and scale-free networks across various social parameters. Both network types exhibit an order-disorder phase transition for all values of \( s \in (0, 1] \). In the model with independence, the system undergoes a continuous phase transition when \( 1 < q \leq 5 \) and a discontinuous phase transition for \( q > 5 \). In contrast, the model with anticonformity consistently shows a continuous phase transition for all \( q > 1 \). The critical point of the model is significantly affected by the type of social behavior, with the critical point in the independence model being lower than in the anticonformity model. This suggests that the independence model is more likely to reach a disordered state, while the anticonformity model tends to maintain order more effectively.

Secondly, the influence of the agent size \( q \) with a fixed \( s \) on the critical point reveals distinct patterns between the independent model and the anticonformity model. In the independence model, the critical point decreases as \( q \) increases, whereas in the anticonformity model, the critical point increases with increasing \( q \). Furthermore, for a fixed \( q \), particularly in the model on a complete graph, different behaviors are observed regarding the effect of the probability \( p \) on the parameter \( s \). As reported in Ref.~\cite{civitarese2021external}, in the independence model, the transition in \( s \) is not smooth for large \( q \) but becomes smooth for small \( q \) at low \( p \). In contrast, in the anticonformity model, the transition in \( s \) remains smooth across all values of \( q \). This difference is closely linked to the discontinuous phase transition characteristic of the independence model and the continuous phase transition observed in the anticonformity model.

Thirdly, the scaling behavior concerning changes in the parameter \( s \) reveals consistent statistical properties for both the model on the complete graph and the scale-free network. The scaling plots, applicable to both network types, show that all data for \( s \) in terms of magnetization \( m \), Binder cumulant \( U \), and susceptibility \( \chi \) collapse effectively, indicating robust scaling behavior. This property holds across different network topologies, including 2D and 3D lattice networks (results not shown in this paper), further underscoring the generality of the scaling behavior. Additionally, finite-size scaling analysis confirms that the critical exponents for both models on the complete graph and the scale-free network are identical, suggesting that the models belong to the same universality class as the mean-field Ising universality class.

Finally, from a social perspective, this study indicates that systems involving independent entities are more prone to polarization than those with anticonformist entities. In other words, systems with anticonformist agents are more likely to maintain consensus, where majority and minority opinions coexist. This finding enhances our understanding of how nonconformity behaviors influence opinion dynamics within complex systems and highlights the role of skepticism in driving social phase transitions. Furthermore, this study provides valuable insights into how skepticism shapes societal consensus and polarization, offering a deeper understanding of the factors that govern these critical social dynamics.
%%%%%%%%%%%%%%%%%%%%%%%%%%%%%%%%%%%%%%%%%%%%%%%%%%%%%%%%%%%%%%%%%%%%%%%%%%%%%%%%
\section*{CRediT authorship contribution statement}
%------------------------------------------------------------------------------
\textbf{R.~A.~NQZ:} Conceptualization, Formal analysis, Writing, Validation, funding acquisition, Supervision, Review \& editing. \textbf{R.~Muslim:} Main contributor, Conceptualization,  Methodology, Writing, Software, Formal analysis, Validation, Visualisation, Review \& editing. \textbf{H.~Lugo H.:} Writing, Software, Formal analysis, Visualisation. \textbf{F.~Nugroho \& I.~S.~Alam:} Formal analysis, Validation, Review \& editing. \textbf{M.~A.~Khalif:} Software, Formal analysis, Validation. All authors have read, reviewed, and approved the publication of this paper.
% %%%%%%%%%%%%%%%%%%%%%%%%%%%%%%%%%%%%%%%%%%%%%%%%%%%%%%%%%%%%%%%%%%%%%%%%%%%%%%%%
\section*{Declaration of Interests}
%-------------------------------------------------------------------------------
The authors declare that they have no known competing financial interests or personal relationships that could have appeared to influence the work reported in this paper.
% %%%%%%%%%%%%%%%%%%%%%%%%%%%%%%%%%%%%%%%%%%%%%%%%%%%%%%%%%%%%%%%%%%%%%%%%%%%%%%%%
\section*{Acknowledgments}
%-------------------------------------------------------------------------------
The authors express their gratitude to the Ministry of Education, Culture, Research, and Technology of Indonesia (Kemendikbudristek) for its financial support through the Regular Fundamental Research Scheme, under contract number 2661/UN1/DITLIT/PT.01.03/2024. \textbf{R.~Muslim} was supported by the Young Scientist Training Program at the Asia Pacific Center for Theoretical Physics (APCTP), funded by the Science and Technology Promotion Fund and Lottery Fund of the Korean Government, as well as by the Management Talent Program of the National Research and Innovation Agency of Indonesia (BRIN).

\appendix
\section{\label{app:A} Critical probability of the model}
The critical probability of the model can also be determined directly from the stationary probability, as shown in Eqs.~\eqref{eq:Eq20} and \eqref{eq:Eq22}. For the model with independence, the stationary probability is expressed as follows: 
\begin{equation}\label{A.1}
    p = \left[\dfrac{1}{2}\left(s\left(1-2\,k_{st}\right)\left[k_{st}\left(1-k_{st}\right)^q-k_{st}^q\left(1-k_{st}\right)\right]^{-1}\right)+1\right]^{-1}.
\end{equation}
The critical point of the model can be obtained by examining the behavior of \( p \) around \( k_{st} = 1/2 \). For this analysis, let \( k_{st} = 1/2 + \delta \), where \( \delta \ll 1 \). By performing a Taylor expansion of the relevant terms and neglecting higher-order terms of \( \delta \), the expression \( \left[k_{st}\left(1-k_{st}\right)^q - k_{st}^q\left(1-k_{st}\right)\right] \) simplifies to:
\begin{align}
    k_{st}(1 - k_{st})^q - k_{st}^q(1 - k_{st}) & \approx \left( \frac{1}{2^{q+1}} + \frac{1-q}{2^q} \right) - \left( \frac{1}{2^{q+1}} - \frac{1-q}{2^q} \right) \nonumber \\
    & = 2^{1-q} \left(1-q\right) \delta. \label{eq:A.2}
\end{align}
Thus, the critical probability \( p_c \) of the model can be expressed as:
\begin{align}
    p_c &= \left[\dfrac{1}{2}\left( \dfrac{-2s\delta}{2^{1-q} \left(1-q\right) \delta}\right)+1\right]^{-1} =\dfrac{\left(q - 1\right)2^{1-q}}{\left(q - 1\right)2^{1-q} + s}
\end{align}
which is identical to Eq.~\eqref{eq:pc_indep}.

The stationary probability for the model with anticonformity is given by:
\begin{equation} \label{eq:A.4}
p = \left[s\left(\dfrac{k_{st}^{q+1}-\left(1-k_{st}\right)^{q+1}}{k_{st}^q\left(1-k_{st}\right)-k_{st}\left(1-k_{st}\right)^q}\right)+1\right]^{-1}.
\end{equation}
By performing a Taylor expansion, the term \( k_{st}^{q+1}-\left(1-k_{st}\right)^{q+1} \) becomes:
\begin{align}
    k_{st}^{q+1}-\left(1-k_{st}\right)^{q+1} \approx & \left(\dfrac{1}{2}\right)^{q+1} + \left(q+1\right) \left(\dfrac{1}{2}\right)^{q} \delta - \left(\dfrac{1}{2}\right)^{q+1} \nonumber \\
    & + \left(q+1\right) \left(\dfrac{1}{2}\right)^{q} \delta \nonumber\\
    =&\,2^{1-q} \left(q+1\right) \delta,
\end{align}
Using the result from Eq.~\eqref{eq:A.2}, the critical point of the model is:
\begin{align}
    p_c  = \left[s\left(\dfrac{2^{1-q} \left(q+1\right) \delta}{2^{1-q} \left(q-1\right) \delta}\right)+1\right]^{-1} = \dfrac{q - 1}{q-1 + \left(q+1\right)s},
\end{align}
which is identical to Eq.~\eqref{eq:pc_anti}.

\section{Critical exponents of the model with anticonformity for $q = 6$}
\begin{figure}[h]
    \centering
    \includegraphics[width=0.9\linewidth]{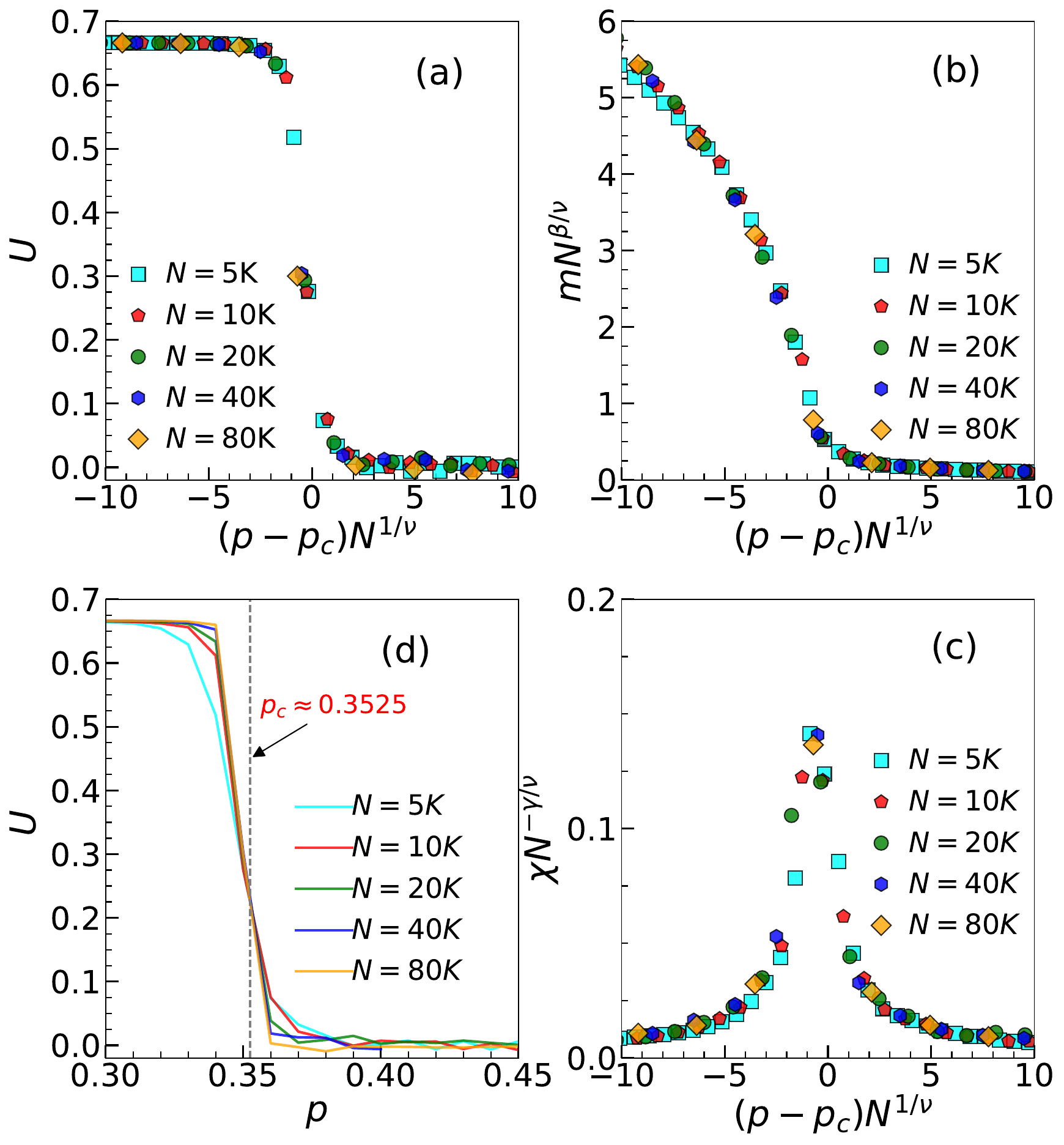}
    \caption{Numerical simulation results of the model with anticonformity for \( q = 6 \) and \( s = 0.8 \) on a BA network with an average degree \( \langle \kappa \rangle = 12 \). The model exhibits a continuous phase transition with a critical point at \( p_c \approx 0.3525 \). FSS analysis demonstrates excellent data collapse near the critical point for critical exponents \( \beta \approx 0.5 \), \( \nu \approx 2.0 \), and \( \gamma \approx 1.0 \), confirming that the model belongs to the same universality class as the mean-field Ising model.}
    \label{fig:scale-free_q6} 
\end{figure}

The numerical results for the model with anticonformity for \( q = 6 \) and \( s = 0.8 \), defined on a Barabási-Albert network with an average degree \( \langle \kappa \rangle = 12 \), are presented in Fig.~\ref{fig:scale-free_q6}. Using finite-size scaling (FSS) analysis, the critical point is identified at \( p_c \approx 0.3525 \). The best-fitting critical exponents are determined as \( \beta \approx 0.5 \), \( \nu \approx 2.0 \), and \( \gamma \approx 1.0 \). These findings confirm that the model belongs to the same universality class as the mean-field Ising model.
\bibliographystyle{elsarticle-num} 
% \bibliography{cas-refs}

\end{document}